\documentclass{article}

\usepackage{amssymb,amsmath,bm}
\usepackage{mathrsfs}  % calligraphic letters
\usepackage{natbib}
\usepackage{fullpage}
\usepackage{amsfonts,amsbsy}
\usepackage{graphicx}
\let\ssection=\section
\renewcommand{\section}{\setcounter{equation}{0}\ssection}

\newcommand{\divv}{\mathop{\mathrm{div}}\nolimits}
\newcommand{\curl}{\mathop{\mathrm{curl}}\nolimits}

\newcommand{\trace}{\mathop{\mathrm{tr}}\nolimits}

\newcommand{\eps}{\varepsilon}

\newcommand{\Ac}{\mathcal{A}}

\newcommand{\Kc}{\mathcal{K}}

\newcommand{\Mc}{\mathcal{M}}

\newcommand{\Sc}{\mathcal{S}}
\newcommand{\Vc}{\mathcal{V}}

\def\realab{\iota}

\newcommand{\Rbb}{\mathbb{R}}

\newcommand{\emf}{\mathcal{E}}
\newcommand{\lie}{\mathcal{L}}

\newcommand{\ip}{\lrcorner}

\newcommand{\omegav}{\boldsymbol{\omega}}

\newcommand{\av}{\boldsymbol{a}}
\newcommand{\bv}{\boldsymbol{b}}
\newcommand{\fv}{\boldsymbol{f}}
\newcommand{\jv}{\boldsymbol{j}}

\newcommand{\uv}{\boldsymbol{u}}
\newcommand{\xv}{\boldsymbol{x}}
\newcommand{\pv}{\boldsymbol{p}}

\newcommand{\bflux}{B}
\newcommand{\bvec}{b}
\newcommand{\bflat}{\beta}

\newcommand{\alphab}{\bar{\alpha}}
\newcommand{\phib}{\bar{\phi}}
\newcommand{\zetab}{\bar{\zeta}}
\newcommand{\ub}{\bar{u}}

\newcommand{\bvecb}{\bar{\bvec}}
\newcommand{\bfluxb}{\bar{\bflux}}

\newcommand{\gh}{\tilde{g}}
\newcommand{\muh}{\tilde{\mu}}
\newcommand{\vh}{\tilde{v}}
\newcommand{\wh}{\tilde{w}}

\newcommand{\nuh}{\tilde{\nu}}

\newcommand{\sigmah}{\tilde{\sigma}}

\newcommand{\deltah}{\tilde{\delta}}

\newcommand{\nablah}{\tilde{\nabla}}

\newcommand{\emfh}{\tilde{\emf}}

\newcommand{\pih}{\tilde{\pi}}

\newcommand{\bvech}{\tilde{\bvec}}
\newcommand{\bflath}{\tilde{\bflat}}
\newcommand{\bfluxh}{\tilde{\bflux}}

\newcommand{\starh}{\tilde{\star}}

\newcommand{\varsigmah}{\tilde{\varsigma}}
\newcommand{\Gammah}{\tilde{\Gamma}}

\newcommand{\dxh}{\tilde{\dd x}}
\newcommand{\dyh}{\tilde{\dd y}}
\newcommand{\dzh}{\tilde{\dd z}}

\newcommand{\pxh}{\tilde{\upartial_x}}
\newcommand{\pyh}{\tilde{\upartial_y}}
\newcommand{\pzh}{\tilde{\upartial_z}}

\newcommand{\pseudop}{\mathsf{p}}
\newcommand{\pseudoh}{\mathsf{h}}

\newcommand{\barL}[1]{\overline{#1}^\mathrm{L}}

\newcommand{\cdotv}{\,{\bm \cdot}\,}
\newcommand{\nablav}{{\bm \nabla}}
\newcommand{\dd}{{\mathrm d}}
\newcommand{\upartial}{\partial}

\def\U{\mathscr{U}}
\def\B{\mathscr{B}}

\begin{document}

\title{A geometric look at MHD and the Braginsky dynamo}
\author{Andrew D. Gilbert$^1$ \& Jacques Vanneste$^2$}
\date{
\normalsize{$^{1}$Department of Mathematics, College of Engineering, Mathematics and Physical Sciences, University of Exeter, Exeter EX4 4QF, UK \\
$^{2}$School of Mathematics and Maxwell Institute for Mathematical Sciences, University of Edinburgh, Edinburgh EH9 3FD, UK} \bigskip \\ \today}

\maketitle

\begin{abstract}

\noindent
This paper considers magnetohydrodynamics (MHD) and some of its applications from the perspective of differential geometry,
considering the dynamics of an ideal fluid flow and magnetic field on a general three-dimensional manifold, equipped with a metric and an induced volume form. The benefit of this level of abstraction is that it clarifies basic aspects of fluid dynamics such as how certain quantities are transported, how they transform under the action of mappings (for example the flow map between Lagrangian labels and Eulerian positions), how conservation laws arise, and the origin of certain approximations that preserve the mathematical structure of classical mechanics.

First, the governing equations for ideal MHD are derived in a general setting by means of an action principle, and making use of Lie derivatives. The way in which these equations transform under a pull back,  by the map taking the position of a fluid parcel to a background location, is detailed. This is then used to parameterise Alfv\'en waves using concepts of pseudomomentum and pseudofield, in parallel with the development of Generalised Lagrangian Mean theory in hydrodynamics. Finally non-ideal MHD is considered with a sketch of the development of the Braginsky $\alpha\omega$-dynamo in a general setting. Expressions for the $\alpha$-tensor are obtained, including a novel geometric  formulation in terms of connection coefficients, and related to formulae found elsewhere in the literature. 

\end{abstract}

\section{Introduction} 

In this paper we discuss some aspects of magnetohydrodynamics (MHD) from a geometric perspective: we consider an ideal fluid flow and magnetic field occupying a three-dimensional manifold $\Mc$ equipped with a metric $g$ and induced volume form $\mu$. We discuss several topics where this level of abstraction is useful in understanding mathematical structure, approximations and phenomena, even  when we may ultimately be operating in everyday three-dimensional space. Taking a geometric perspective on fluid dynamics was pioneered in the seminal paper \cite{Ar66}; this and related works are reviewed in the book \cite{ArKh98}. Building on the mathematical foundations of classical mechanics, Arnold showed that an incompressible ideal fluid flow may be considered as a geodesic on a manifold known as SDiff($\Mc$), the space of all volume-preserving diffeomorphisms of the space $\Mc$, and that this space has an underlying Lie group structure. Here SDiff($\Mc$) has a metric that comes from the metric $g$ on $\Mc$, and this is needed to define geodesics: the shortest path taken by a point in this big space SDiff($\Mc$) gives the flow map, say $\phi_t$, that moves all the Lagrangian fluid parcels for the actual flow in $\Mc$. Arnold's geometric perspective on a classical topic led to many new and varied results in theoretical fluid dynamics, for example in stability theory \citep[e.g.][]{HoMaRaWe85}, in  Hamiltonian mechanics, understanding the origin of conservation laws and determining systematic approximate or reduced fluid systems   \citep[e.g.][]{Mo82,McSh87,Sa88,Sh90,Mo98,We18}, and in modelling fluctuations and waves in laminar and turbulent flows \cite[e.g.][]{FoHoTi01,SoRo10,GiVa18}. Often the key advantage of a geometric perspective is that the transformations or approximations used naturally preserve the underlying geometry of fluid motion, in other words the essential structure of classical mechanics, together with its conservation laws. A modern geometric approach also underlies \emph{mimetic} numerical methods \citep[e.g.][]{QiGuTa09,LeSh12,ThCo15,KrTaGr16,PaGe17}, which are designed to respect key mathematical properties; an example is that a discrete approximation to vector calculus operators should preserve the fundamental identity that $\divv \curl \fv = 0$, or in a more general setting that $\dd^2 \gamma = 0 $ for the exterior derivative $\dd$ of a differential form $\gamma$. Failing to preserve such identities easily leads to  spurious physical effects whose origin is purely numerical. 

While there has been more study of geometric methods applied to hydrodynamics than to MHD \citep[reviewed in the book][]{We18}, many key ideas in fact arose in parallel with (or following) similar developments in the latter field. One notion familiar to anyone dealing with a magnetic field $\bv$ (or vorticity $\omegav$) is that ideal transport (that is, with zero diffusion) in incompressible flow is achieved by means of the Cauchy solution. 
%This relates a field vector $\bv(\xv)$ attached to a Lagrangian fluid parcel at say $\xv$ that is transported to a point $\xv'$ with the new vector $\bv'(\xv')$ equal to the Jacobian matrix $J$ of the transformation $\psi: \xv \to \xv'$ multiplied into $\bv(\xv)$, 
%
%\begin{equation}
%\bv'(\xv') = J(\xv) \bv(\xv) , \quad J_{ij} = \upartial x'_i /\upartial x_j. 
%\label{eqJb} 
%\end{equation}
%
%\JVcom{Maybe this is pedantic but I think it'd be better to write this in coordinates. Alternative would be: \\
This relates a field vector $\bv(\xv)$ attached to a Lagrangian fluid parcel at  $\xv$ that is transported to a point $\xv'$, to the new vector $\bv'(\xv')$ with components
\begin{equation}
b_i'(\xv') = J_{ij} (\xv) b_j(\xv) , \quad \textrm{where} \ \  J_{ij} = \upartial x'_i /\upartial x_j
\label{eqJb} 
\end{equation}
is the Jacobian matrix $J$ of the transformation $\psi: \xv \mapsto \xv'$.
%}

In geometric parlance this is the \emph{push forward} of the field under the map $\psi$, and is a standard way of looking at the transport of magnetic field in ideal or near-ideal MHD or dynamo theory. What is less intuitive perhaps, is that some quantities are not transported in this way, as a (contravariant) vector field, but as a 1-form field (or covariant vector field), an example of this being momentum. In a key paper, \cite{So72} showed that the natural way of tranporting momentum $\pv$ from place to place is using the inverse transpose of the same Jacobian, 
%%
%\begin{equation}
%\pv'(\JV{\xv'}) = K(\JV{\xv}) \pv(\JV{\xv}) , \quad K = [J^{-1}]^{\mathrm{T}}.  
%\label{eqKp} 
%\end{equation}
%%
%\JVcom{Alternative: \\
\begin{equation}
p'_i(\xv') = K_{ij}(\xv) p_j(\xv) , \quad K = \left(J^{-1}\right)\!{}^{\mathrm{T}}.  
\label{eqKp} 
\end{equation}
%}
In the language of differential geometry this is a push forward of the momentum $\pv$ considered as a 1-form, under the map $\psi$.

Thus, there are two types of transport relevant to everyday vector fields in incompressible flow, that given by (\ref{eqJb}) and that given by (\ref{eqKp}). If we have quantities transported in this way then the inner product $\bv \cdotv \pv$ is conserved. As an example, if the map $\psi$ is the flow map $\phi_t$, moving Lagrangian parcels from their positions at time $t=0$ to those at time $t$, then magnetic field $\bv$ and vorticity $\omegav$ evolve according to (\ref{eqJb}), and a scalar gradient $\nablav \chi$ and (in a suitable gauge) the magnetic vector potential $\av$ evolve according to (\ref{eqKp}). We then have immediately that $\omegav\cdotv\nablav \chi$ and $\bv\cdotv\av$ are conserved on fluid parcels. These are respectively Ertel's potential vorticity \citep{Er42} and the magnetic helicity density \citep[see][]{MoDo19}. 

The picture is more complicated for the momentum $\pv$ because this can be redistributed by the pressure field, so that the transport of momentum does not occur via (\ref{eqKp}) under the flow map $\phi_t$. However this equation does give the means of moving momentum from point to point in any kind of Lagrangian averaging. This was realised in Soward's (1972) paper, which built on earlier work of  \cite{FrRo60} and \cite{Ec63}, and was developed as Generalised Lagrangian Mean (GLM) theory in the papers \cite{AnMc78a,AnMc78b}; see the book \cite{Bu09} for further developments. The key idea here is that it is often beneficial to take averages of fluid dynamical equations not at a fixed location in a family of flow realisations -- an Eulerian average -- but to take a Lagrangian average, over the locations of a single Lagrangian parcel. In this case it is necessary to move vectors from place to place in the flow, and the transformation rules (\ref{eqJb}) and (\ref{eqKp}) come into play. If quantities are moved in this way (rather than parallel transport, say), remarkable properties of the resulting Lagrangian averages emerge; in particular the structure of the Euler equation is preserved to the extent that Kelvin's circulation theorem continues to apply; for recent developments see \cite{Ho02b}, \cite{SoRo10}, \cite{GiVa18}  and the book \cite{We18}. 

Although all this machinery works beautifully and is well established in the literature, together with applications, we argue that it is only by stepping back and considering some of these fluid dynamical systems in a more abstract setting that it is clear why these methods work, what is the origin of various transformation and conservation laws, and where choices can and cannot be made. For example working in Euclidean space $\Rbb^3$ with Cartesian coordinates $(x,y,z)$ and metric $g = \dd x^2 + \dd y^2 + \dd z^2$, it is too easy to switch between vector fields such as the velocity $\uv$ and 1-form fields such as the momentum $\pv$, which have the same components (up to the factor of density $\rho$) even though they are very different objects in terms of their properties under transport. Working on a general manifold $\Mc$ with a metric $g$ and an induced volume form $\mu$ forces one to establish  what type of object one is dealing with at the outset, and with this, theory can become easier and clearer, drawing on well-established results in differential geometry. The advantage of the use of differential geometry and particularly Lie derivatives has been stressed by several authors in MHD, including \cite{TuYa93}, \cite{MaSh01}, \cite{Ho02b}, \cite{RoSo06a,RoSo06b}, \cite{Ar13a,Ar13b}, \cite{SoRo14} and \cite{We18}. Even working in $\Rbb^3$, the use of a formulation based on a general metric $g$ can avoid complications switching from Cartesian coordinates to, say, cylindrical polars, and is well-nigh essential if non-orthogonal coordinates are employed, for example when a coordinate labels surfaces of constant buoyancy in a geophysical fluid dynamical setting \citep{GiVa20b}. 

In the present paper, we discuss three related topics in which we place MHD in an abstract geometric setting. In section 2 we derive the MHD equations from an action principle: this starting point is useful as it forces us to identify the different types of objects that can be used to describe the flow field (velocity $u$, momentum $\rho\nu$) and magnetic field (flux 2-form $B$, vector field $b$ and 1-form field $\beta$), how they are related using the metric and volume form, and how they appear in the equations of motion. Compressible fluid flow is allowed, and the magnetic pressure term emerges in the calculation. 
The section is standard material but takes a geometric approach to recast the classical derivation of \cite{Ne62}, reviewed in \cite{VaFe67}, \cite{Mo98} and \cite{We18}. We also discuss the conservation of different types of helicity within the geometical framework \citep{ArKh98,Fu08,WeDaMcHuZa14a,WeDaMcHuZa14b}.
With the equations established in geometric form, in section 3 we discuss how they transform under a pull back (or push forward) by a mapping. This gives rise to the notion of pseudomomentum and pseudo (magnetic) field, and these are calculated for finite amplitude Alfv\'en waves on a uniform magnetic field threading an incompressible fluid \citep{AlFa50}. We also consider cross helicity under Lagrangian averaging \citep{Ho02a}.  

In section 4 we discuss transformation of the induction equation under a mapping and sketch its use in the \cite{Br64a,Br64b} dynamo: here one starts with a background flow that is not a dynamo but gives an $\omega$-effect, stretching out field that is transverse to streamlines. Then waves are included by means of a Lagrangian map: through the diffusion term, these can generate an $\alpha$-effect and close the dynamo cycle to give an $\alpha\omega$-dynamo. We note that the use of mappings and transport of field is well-established following \cite{So72}, with further developments in \cite{RoSo06a,RoSo06b} and a comprehensive treatment given recently in \cite{SoRo14}. Although these papers use many concepts from differential geometry implicitly, the goal of the present paper is to make the machinery explicit, and so both simplify and unify previous methods and results: to our knowledge the perspective that we aim to convey is novel and we also obtain a new description of the $\alpha$-effect in terms of a connection tensor.  The present paper is self-contained, but the framework follows on from \cite{GiVa18}; a companion paper \cite{GiVa20a} studies how fluid systems can be written in a conservation form, including MHD and shallow water MHD. 
 
\section{Ideal MHD from an action principle}

\subsection{Usual MHD equations} 

For reference purposes, we begin by setting out the equations for MHD in usual Euclidean space $\Rbb^3$ using standard notation, namely, 
\begin{align}
\rho ( \upartial_t \uv &  + \omegav \times \uv + \tfrac{1}{2} \nablav  |\uv|^2 ) +  \nablav p = \jv \times \bv ,
\label{eqeverydaymom}  \\
 \upartial_t \bv & = \nablav \times (\uv \times \bv)   -\eta  \nablav \times \jv,
 \label{eqeverydayind}  \\
\omegav&  = \nablav\times \uv , \quad \jv  = \nablav \times \bv , \quad \nablav \cdotv \bv = 0 . 
\label{eqeverydaymisc} 
\end{align}
Here we have allowed the field to be non-ideal, with (constant) magnetic diffusivity $\eta$: sections 2 and 3 will consider ideal MHD, $\eta=0$, and section 4 dynamos with $\eta>0$.  We have not introduced viscosity; see \cite{GiRiTh14} and \cite{GiVa20a} for discussion of how the divergence of a stress tensor is taken in a geometric setting. For convenience the magnetic permeability $\mu_0$ is scaled out, the true magnetic field is in fact $\mu_0^{1/2} \bv$.  
%
% see HKM book page 316, also for density. 
%
We allow either incompressible flow with $\nablav \cdotv \uv = 0$ or ideal compressible flow, in which case an equation of state $p = p (\rho, s)$ is also needed, with $s$ as the entropy and $\rho$ the density governed by 
\begin{align}
 \upartial_t s & + \uv \cdotv \nablav s = 0,   \\
 \upartial_t \rho & + \nablav \cdotv (\rho \uv) = 0 . 
\end{align}

\subsection{Geometric setting and transport} 

Having set out the equations in ordinary three-dimensional space, we now place these in an abstract setting, and will use lighter face quantities for all the various physical fields to stress this. Our fluid domain is now an orientable three-dimensional manifold $\Mc$, with or without a boundary $\upartial\Mc$; examples include all of $\Rbb^3$ or a solid sphere. For simplicity we will avoid discussion of manifolds with `holes' in them, for example a spherical shell; in other words we restrict to manifolds $\Mc$ where any curve or closed surface can be contracted to a point. In ideal fluid mechanics any flow or magnetic vector field is taken parallel to the boundary,  $u , b \parallel \upartial \Mc$. 

We assume that the reader has knowledge of the fundamentals of differential geometry, as given by, for example, \cite{HaEl73},  \cite{Sc80}, \cite{Fr97} or \cite{BeFr17}. In particular we make use of vectors, differential forms, the Lie derivative $\lie$, inner product $\ip$, exterior derivative $\dd$, the Hodge star operator $\star$, and the musical raising and lowering operators $\sharp$ and $\flat$.  To discuss fluid dynamics we need $\Mc$ to be equipped with a metric $g$ and induced volume form $\mu$, and occasionally it is useful to refer to the corresponding covariant derivative $\nabla$. We make frequent use, often without comment, of Cartan's formula: for any differential form $\gamma$, 
\begin{equation}
\lie_u \gamma = \dd (u \ip \gamma) + u \ip \dd \gamma. 
\label{eqCartan} 
\end{equation}

Our first concern is the magnetic field, whose most fundamental property perhaps is the absence of magnetic monopoles, so that the integral of magnetic flux over any closed surface $\Sc$ vanishes. From the geometric viewpoint this means that the magnetic field is most naturally represented by the \emph{magnetic flux 2-form} $\bflux $ \citep{TuYa93,Fr97}, which is required to be closed, $\dd\bflux  = 0$. Then, as any closed surface $\Sc$ in $\Mc$ bounds a volume $\Vc$ with $\Sc = \upartial \Vc$, we have 
\begin{equation}
\int_{\Sc} \bflux  = \int_{\Vc} \dd\bflux  = 0 , 
\end{equation}
using the generalised Stokes theorem (corresponding to the usual divergence theorem in this instance). The focus on the magnetic flux $\bflux$ is natural in a geometric setting since 2-form fields are naturally integrated over surfaces. In ideal MHD, magnetic field is transported in the fluid flow, as per Alfv\'en's theorem, and in the general setting this corresponds to requiring that $\bflux $ be Lie dragged in the fluid flow  \citep{TuYa93}, 
\begin{equation}
\upartial_t \bflux  + \lie_u \,\bflux  = 0 . 
\label{eqlieB}
\end{equation}
This preserves the  condition $\dd\bflux =0$ since $\dd$ commutes with Lie and time derivatives. 
%\AG{This may need changing:} 
%\AGadd{(Note that $\bflux$ and $\bflat$ were called $B$ and $\beta$ respectively in \cite{GiVa19a}.) or (We aim to use greek letters for forms but $B$ is an exception --- we have run out of helpful letters for magnetic quantities.) }

We are often more used to thinking of magnetic field as a \emph{magnetic vector field} $b$ rather than a 2-form field $\bflux $, and $\bvec$ is easily defined using the volume 3-form via $\bvec\ip \mu = \bflux $. In this case it follows from (\ref{eqlieB}) that $\bvec$ is transported according to 
\begin{equation}
\upartial_t \bvec+ \lie_u \,  \bvec  + \bvec\,  \divv u = 0 , 
\label{eqlieb}
\end{equation}
where $\divv u$ is the divergence of the flow field $u$ given by 
\begin{equation}
\lie_u \, \mu  = \dd(u\ip \mu) = \mu \divv u. 
\label{eqliemu}
\end{equation}
This transport of $\bvec$ takes into account the action of compressible flow on the field through the $\bvec \,\divv u$ term, and for this reason $\bvec$ is sometimes referred to as a `tensor of weight $-1$' \citep{RoSo06a}. The solenoidal property of $b$, namely $\divv b =0 $, follows from $\dd B=0$. 

Other transported quantities include the entropy $s$ which is a scalar field obeying
\begin{equation}
\upartial_t s + \lie_u\,  s = 0 , 
\label{eqlies}
\end{equation}
and density $\rho$. Again, more fundamental than density itself is the \emph{mass 3-form} $m $, in that integration of $m$ over a volume gives the mass contained therein. %, and  3-forms can be integrated over volumes. 
The mass form is again Lie dragged via 
\begin{equation}
\upartial_t m + \lie_u \, m = 0 , 
\label{eqliem}
\end{equation}
giving mass conservation, while the density $\rho$, now defined by $m = \rho \mu$, is a scalar of weight $-1$ obeying
\begin{equation}
\upartial_t \rho + \lie_u \, \rho + \rho \, \divv u   = 0 . 
\label{eqlierho}
\end{equation}
In short, in an ideal setting the fundamental quantities of magnetic flux $\bflux$,  entropy $s$ and mass $m$ are all Lie dragged in the flow field in the same way via (\ref{eqlieB}), (\ref{eqlies}) and (\ref{eqliem}), while the derived quantities of magnetic vector field $\bvec$ and density $\rho$ obey the slightly more complex equations (\ref{eqlieb}) and (\ref{eqlierho}), bringing in the divergence of the flow field $u$. This latter effect can also be absorbed by dividing $b$ by $\rho$; see \cite{Ar13a} for a development along these lines, and see \cite{TuYa93} for discussion of Lie-dragging and how ideal invariants are systematically formulated within this framework.

\subsection{Ideal MHD from an action principle} 

We now derive the equations for the flow $u$ starting with an action principle \citep{Ne62,Mo98,We18}. We follow the geometrical discussion in \cite{GiVa18} 
%(which can be consulted for more background)
with an additional magnetic field; see \cite{HoMaRa} for an equivalent derivation in the more general framework of Euler--Poincar\'e systems, and \cite{ArKh98} for the treatment of incompressibility case. For this we need the Lagrangian flow map $\phi \equiv \phi_t$ (we drop the subscript $t$) a time-dependent diffeomorphism of $\Mc$ which takes fluid elements from their positions $x$ at time $t=0$ to their current position $\phi(x,t)$. This map is assumed to be invertible and differentiable as much as is needed. In terms of $\phi$ the flow $u$, which is a time-dependent vector field on $\Mc$, is given by 
\begin{equation}
u  = \dot{\phi} \circ \phi^{-1}. 
\label{equphi}
\end{equation}
The kinetic energy of the fluid may then be expressed in terms of an integral over the initial particle locations (i.e.\ at $t=0$), used as Lagrangian labels,  by 
\begin{equation}
\Kc = \int_{\Mc} \tfrac{1}{2} g (\dot{\phi}, \dot{\phi}) \, m_0 , 
\end{equation}
where $m_0 = \rho_0 \mu $ is the initial mass distribution, or in terms of the current positions and current mass distribution $m$ (i.e.\ at a general time $t$) by 
\begin{equation}
\Kc = \int_{\Mc} \tfrac{1}{2} g (u, u)\,  m .
\end{equation}
The full action,  in terms of current positions, is then
\begin{equation}
\Ac[\phi] = \int \dd t \int_{\Mc} \big[ \, \tfrac{1}{2} g(u, u)  m -  \rho e(\rho,s) \mu -   \tfrac{1}{2} g(\bvec , \bvec ) \mu \,\big] .
\label{eqaction} 
\end{equation}
Here the fields $m$, $s$, $\bflux$, $\rho$ and $\bvec$ depend on the flow map since they are obtained from their initial values by the push forwards, 
%
%where the various fields have been evolved in the flow $u$, from their initial distributions. This  corresponds to their being pushed forwards by the flow map $\phi$, with 
%
\begin{equation}
 m = \phi_* m_0, \quad 
s = \phi_* s_0, \quad  
\bflux  = \phi_* \bflux _0 , 
\label{eqtransportfields}
\end{equation}
and as above $m = \rho\mu$ and $\bflux  = \bvec \ip \mu$. For magnetic field this corresponds to the Cauchy solution. In the action, $e(\rho,s)$ is the internal energy per unit mass, and it is notable that this and the kinetic energy are weighted with mass $m= \rho \mu$, whereas the magnetic energy involves the vector field $\bvec$ rather than the flux 2-form $\bflux $, and is weighted with volume $\mu$. 

Using Hamilton's principle, we require that the action is stationary under variations in the paths of the fluid particles over some time interval. We achieve this in the present framework by replacing the flow map $\phi$ in the above by the perturbed flow map $\phi_\eps = \psi_\eps \circ \phi$. Here $\psi_\eps$ is a family of diffeomorphisms of $\Mc$ dependent on time and on a scalar parameter $\eps$, equal to the identity for $\eps=0$. We require that the action be stationary with respect to such variations, 
\begin{equation}
\frac{\dd}{\dd\eps} \Big|_{\eps=0}  \, \Ac [ \phi_\eps ] = 0 . 
\label{eqactionvary}
\end{equation}
If we fix time $t$ and vary $\eps$ around $\eps=0$ the family of maps $\psi_\eps$ gives a vector field $w$, formally defined by 
\begin{equation}
w =  \left. \frac{\dd \psi_\eps}{\dd\eps}  \circ \psi_{\eps}^{-1} \right|_{\eps=0}, 
\label{eqwheredef} 
\end{equation}
and this  field is parallel to the boundary of $\Mc$. On the other hand if we fix $\eps$ and vary $t$ we obtain the flow velocity under $\phi_\eps$ as in (\ref{equphi})  with 
\begin{equation}
u_\eps = \dot{\phi}_\eps \circ \phi_\eps^{-1}. 
\label{equepsdef} 
\end{equation}
Key is the relationship between $u_{\eps}$ and $w$, how a variation in the map affects the resulting particle velocities, and this is 
\begin{align} 
& \left. \frac{\dd}{\dd\eps} \right|_{\eps=0} u_\eps = \upartial_t w + \lie_u\,  w = \upartial_t w - \lie_w \, u .
\label{eqlieuw}
\end{align}
This key identity is easily shown using coordinates (or more abstractly in \cite{GiVa18}, appendix B) by writing from (\ref{eqwheredef}, \ref{equepsdef})
\begin{equation}
u_\eps(\phi_\eps(x,t),t) = \upartial_t \phi_\eps(x,t), \quad
w_\eps(\phi_\eps(x,t),t) = \upartial_\eps \phi_\eps(x,t), 
\end{equation}
and equating $\upartial_t \upartial_\eps \phi _\eps = \upartial_\eps\upartial_t \phi_\eps$ to give 
\begin{equation}
\upartial_\eps u^i_\eps + (\upartial_j u_\eps^i )( \upartial_\eps \phi^j_\eps )  = \upartial_t w^i_\eps + (\upartial_j w^i_\eps)(  \upartial_t \phi^j_\eps ), 
\end{equation}
or
\begin{equation}
\upartial_\eps u^i_\eps + (\upartial_j u_\eps^i )w_\eps^j   = \upartial_t w^i_\eps + (\upartial_j w^i_\eps)u_\eps^j  , 
\end{equation}
which, evaluated at $\eps=0$, amounts to (\ref{eqlieuw}). 

For other quantities in the action, we have from the earlier transport equations (\ref{eqlieB}--\ref{eqlierho}) that 
\begin{align} 
& \left. \frac{\dd}{\dd\eps} \right|_{\eps=0} \, m_\eps  = - \lie_w \,  m  = - \lie_w  (\rho \mu)  = - \divv (\rho w) \mu, \\
& \left. \frac{\dd}{\dd\eps} \right|_{\eps=0} \, \bvec_\eps =  - \lie_w \, \bvec - (\divv w) \bvec  , \\
& \left. \frac{\dd}{\dd\eps} \right|_{\eps=0} \, s_\eps = - \lie_w\,  s , \\
 & \left. \frac{\dd}{\dd\eps} \right|_{\eps=0} \, \rho_\eps = - \divv (\rho w) , 
\end{align}
where the $\eps$ subscript denotes their values under transport  by $\phi_\eps$ from the same initial conditions in (\ref{eqtransportfields}). 

The requirement (\ref{eqactionvary}) that the action $\Ac[\phi_\eps]$ be stationary at $\eps=0$  becomes  
\begin{align}
\int \dd t \int_\textrm{M} \big[ \,
 g(u, \upartial_t w + \lie_u \, w ) \, m 
& - \tfrac{1}{2} g(u, u) \, \lie_w \, m 
%\notag\\
+  (\rho e) _\rho \divv (\rho w) \mu 
+ \rho  e_s ( \lie_w \, s ) \, \mu
\notag\\
& +    g(\bvec , \lie_w \, \bvec+ (\divv w) \bvec) \, \mu 
 \,\big]  = 0 . 
\label{eqactionvary2}
\end{align}
We need to pull out the arbitrary vector field $w$ from this. First we take $w$ to be identically zero outside some time interval, which we do not need to give explicitly, so we can apply integration by parts with respect to time. Secondly we can use integration by parts in space. Let $\gamma$ be any 3-form, then for any vector field parallel to the boundary of $\Mc$ (such as $u$, $w$, $\bvec$) we have 
\begin{equation}
\int_{\Mc} \lie_u \gamma = \int_{\Mc} \dd(u \ip \gamma) = \int_{\upartial\Mc} u \ip \gamma = 0 . 
\end{equation}
Together with the Leibniz rule for Lie derivatives, this means that we can integrate by parts and shift the Lie derivative from a term in a product to the remaining terms, while introducing a minus sign. 

Using these and similar rules we can replace quantities in the above integral by their equivalents. To keep notation light we denote this by $\simeq$. We define the key quantities, the 1-forms, 
\begin{equation}
\nu = g(u, \cdot) = u_{\flat}, \quad
\end{equation}
and  
\begin{equation}
\bflat =  g(\bvec,\cdot)  = \bvec_{\flat} = \star \bflux, 
\end{equation}
where $\star$ is the Hodge star operator. 
%\JVcom{I rearranged the computation below to avoid the form-valued 3-forms -- I kept the original as comment.} 
We then have for the kinetic energy terms, 
\begin{align}
g(u, \upartial_t w  )  m & =  \mu \, \rho \nu \ip  \upartial_t w   \simeq  - w \ip \upartial_t (\rho\nu) \mu = - w \ip \nu  \,(\upartial_t \rho) \,\mu - w \ip (\upartial_t \nu) \, \rho \mu ,  \\
g(u, \lie_u \,w )  m & = \mu\,  \rho \nu \ip \lie_u \,w \simeq  - w \ip \lie_u ( \rho \nu) \mu - \rho w \ip \nu \, \lie_u\,  \mu  \nonumber \\
& =  - w \ip \lie_u ( \rho \nu) \mu - \rho w \ip \nu \,( \divv u)  \,  \mu =  - w \ip (\lie_u \,  \nu ) \rho \mu +  w \ip \nu \, (\upartial_t \rho)  \, \mu  , \label{eqenergyinter}\\
- \tfrac{1}{2} g(u, u) \, \lie_w \,m &   \simeq m \, \lie_w \tfrac{1}{2} g(u, u) =  m w \ip \tfrac{1}{2} \dd g(u, u)  = w \ip  \tfrac{1}{2} \rho \, \dd g(u, u) \mu,  
\end{align}
using (\ref{eqlierho}) for (\ref{eqenergyinter}).
%
%%
%\begin{align}
%g(u, \upartial_t w  )  m & =  \mu \, \rho \nu \ip  \upartial_t w   \simeq  - w \ip \upartial_t (\rho\nu\otimes\mu),  \\
%g(u, \lie_u \,w )  m & = \mu\,  \rho \nu \ip \lie_u \,w \simeq  - w \ip \lie_u ( \rho \nu \otimes \mu) , \\
%- \tfrac{1}{2} g(u, u) \, \lie_w \,m &   \simeq m \, \lie_w \tfrac{1}{2} g(u, u) =  m w \ip \tfrac{1}{2} dg(u, u)  = w \ip  \tfrac{1}{2} \rho \, dg(u, u) \otimes \mu .   
%\end{align}
%%
For the internal energy terms we have 
\begin{align}
 (\rho e) _\rho \divv (\rho w) \mu  + \rho  e_s ( \lie_w \,s ) \, \mu 
 & =  (\rho e) _\rho \, \lie_w (\rho \mu)   + \rho  e_s ( \lie_w \,s ) \, \mu  \notag \\
   &  \simeq  -    \lie_w [ (\rho e) _\rho ] \rho \mu  + \rho  e_s ( \lie_w \,s ) \, \mu  \notag \\
  &   =
 -  w \ip  \dd [ (\rho e) _\rho  ] \rho \mu   +w \ip (\rho  e_s \,\dd s) \mu   \\
 & = w \ip ( - \rho\, \dd h + \rho T \,\dd s)  \mu  \notag \\
 & =-  w \ip dp \,  \mu , 
 \notag
 \end{align}
%
%%
%\begin{align}
% (\rho e) _\rho \divv (\rho w) \mu  + \rho  e_s ( \lie_w \,s ) \, \mu 
% & =  (\rho e) _\rho \lie_w (\rho \mu)   + \rho  e_s ( \lie_w \,s ) \, \mu  \notag \\
%   &  \simeq  -    \lie_w [ (\rho e) _\rho ] \rho \mu  + \rho  e_s ( \lie_w \,s ) \, \mu  \notag \\
%  &   =
% -  w \ip \rho\, d [ (\rho e) _\rho  ] \otimes\mu   +w \ip (\rho  e_s \,ds) \otimes\mu   \\
% & = w \ip ( - \rho\, dh + \rho T \,ds) \otimes \mu  \notag \\
% & =-  w \ip dp \otimes \mu , 
% \notag
% \end{align}
%%
where we have used thermodynamic relations for pressure $p$, temperature $T$ and enthalpy $h$, 
\begin{equation}
h = (\rho e)_\rho = e + p / \rho, \quad T = e_s, \quad \dd h = \rho^{-1} \dd p + T \dd s. 
\end{equation}
Finally, for the magnetic terms we have 
\begin{align}
   g(\bvec , \lie_w \,\bvec) \mu &   =   -  \mu\, \bflat \ip \lie_\bvec\, w \simeq   w \ip (\lie_\bvec\,  \bflat )  \mu + w \ip \beta \, \lie_b \, \mu = w \ip (\lie_\bvec\,  \bflat ) \mu,  \\
   g(\bvec , (\divv w) \bvec) \mu & = g(\bvec,\bvec) \lie_w \,\mu  \simeq
  -  \lie_w [g(\bvec,\bvec)] \mu = - w \ip \dd g(\bvec,\bvec) \mu,
\end{align}
using that  $ \lie_\bvec \,  \mu = \dd (\bvec\ip \mu) = \dd\bflux   = 0$. 
These expressions are inserted into (\ref{eqactionvary2}) which must hold for arbitrary $w$, and from this we obtain the momentum equation in the  form
%These expressions are inserted into (\ref{eqactionvary2}) which must hold for arbitrary $w$, and from this we obtain the momentum equation in the somewhat unwieldy form
%%
%\begin{align}
%\upartial_t (\rho \nu\otimes \mu)  & + \lie_u (\rho \nu \otimes \mu) -   \tfrac{1}{2}\rho  \, dg(u, u) \otimes \mu + dp \otimes \mu   = \lie_\bvec  (\bflat \otimes \mu)  - dg(\bvec,\bvec)\otimes \mu, 
%\label{eqmomentum1} 
%\end{align}
%%
%as it is expressed in terms of the tensor products of 1-forms with the volume 3-form $\mu$. To simplify these we note that $m  = \rho \mu$ satisfies (\ref{eqliem}) and so this quantity can be taken out of the first two terms of (\ref{eqmomentum1}).  Also $ \lie_\bvec  \mu = d (\bvec\ip \mu) = d\bflux   = 0 $ and so $\mu$ can come out of the penultimate term, so expressing all quantities as $\text{(1-form)}\otimes \mu$. Then $\mu$ can be taken out to leave
%%
\begin{align}
\rho [ \upartial_t \nu & + \lie_u\, \nu -  \tfrac{1}{2} \dd g(u, u) ]+ \dd p  = \lie_\bvec  \,\bflat  - \dd g(\bvec,\bvec), 
\label{eqmomentum2a} 
\end{align}
for ideal MHD flow in a general setting. 

\subsection{Discussion} 

Several remarks are in order. First, using Cartan's formula (\ref{eqCartan}) and noting that $g(u,u) = \nu \ip u$, $g(b,b) = b \ip \bflat$, (\ref{eqmomentum2a}) may be expressed as 
\begin{align}
\rho [ \upartial_t \nu & + u \ip \dd  \nu + \tfrac{1}{2} \dd g(u, u) ]+ \dd p  = b\ip \dd \bflat, 
\label{eqmomentum2b} 
\end{align}
which is equivalent to (\ref{eqeverydaymom}) where we identify 
\begin{equation}
\zeta = \dd\nu \quad \textrm{and} \quad j = \dd\bflat
\label{eqvorticitycurrent}
\end{equation}
as the vorticity and current 2-forms. Both of these can be converted to vector fields in the same way that $\bvec \ip \mu = \bflux $ relates the magnetic flux 2-form $\bflux $ to the corresponding vector field $\bvec$. Likewise  equation (\ref{eqlieB}) for evolution of magnetic field may be rewritten as 
\begin{equation}
\upartial_t \bflux  + \dd ( u \ip \bflux ) = 0 , 
\label{eqmomentum2bb} 
\end{equation}
which is analogous to (\ref{eqeverydayind}) for $\eta = 0$, with the exterior derivative playing the role of the curl. 

Secondly, in the momentum equation we have found that terms emerge for the transport, not of the velocity vector $u$, but for the corresponding 1-form \emph{momentum} $\nu = u_\flat$ (strictly the momentum is $\rho \nu$). Although these two are very similar, and in relativity would more-or-less be identified, in the context of geometrical fluid dynamics it is important to keep the distinction between the two quantities. Likewise in determining the Lorentz force term we are driven to introduce the 1-form $\bflat$.  Both $\bvec$ and $\bflat$ are related back to the fundamental magnetic flux 2-form $\bflux$ since $\bvec \ip \mu = \bflux $ and $\bflat = \bvec_\flat $, but it is convenient to distinguish them using different symbols. It is not clear that there is a good name for the magnetic 1-form field $\bflat$ (we have the \emph{magnetic vector field} $b$ and the \emph{magnetic flux 2-form} $\bflux$) but the distinction between the two quantities $\bvec$ and $\bflat$ and their properties has been stressed in \cite{Ho02a} and \cite{RoSo06a}, for reasons  that become apparent in section 3.

Thirdly, note that a vorticity equation can be obtained in terms of Lie derivatives \citep{Ar13a} by dividing (\ref{eqmomentum2a}) by $\rho $ to give
\begin{align}
 \upartial_t \nu & + \lie_u\, \nu -  \tfrac{1}{2} \dd g(u, u) +  \rho^{-1} \dd p  =  \rho^{-1}b\ip \dd\bflat, 
\label{eqmomentum2e} 
\end{align}
then applying $\dd$ to obtain
%We then apply $d$, which commutes with Lie derivatives, and again set the vorticity $\zeta = d\nu$ and current $j = d\bflat$, to give, 
%
\begin{align}
 \upartial_t \zeta & + \lie_u \, \zeta  - \rho^{-2}  \dd\rho\wedge \dd p  = \dd (\rho^{-1}\bvec\ip j)  = \lie_{\bvec/\rho}\, j, 
\label{eqvorticity2} 
\end{align}
using (\ref{eqvorticitycurrent}) and Cartan's formula (\ref{eqCartan}). The quantity $\zeta$ is referred to as the \emph{potential vorticity} by \cite{Ar13a}. This should not be confused with the (Rossby--Ertel) potential vorticity 3-form $\zeta \wedge \dd s$, or the potential vorticity scalar $Q$ defined by $\rho Q \mu = \zeta \wedge \dd s$, both of which are Lie transported by $u$ in the absence of magnetic field.

Perhaps a more familiar route for some readers is to express the momentum and induction equations in terms of the covariant derivative. Here we write the components of $u$ as $u^i$ and those of $\nu$ as $u_i$, etc. Using standard results, in particular that 
\begin{equation}
\lie_u \, \nu = (\nabla_u u)_\flat + \tfrac{1}{2}\dd g(u,u)
\end{equation}
\citep{ArKh98}, the momentum equation (\ref{eqmomentum2a}) may be  written as 
\begin{equation}
\rho \left( \upartial_t u_i  + u^j \nabla_j u_i \right) +    \nabla_i \left(  p  +  \tfrac{1}{2} g_{jk} b^j b^k \right) = b^j \nabla_j b_i .  
\label{eqmomentum2c} 
\end{equation}
In this formulation it is interesting to see the magnetic pressure term $ \tfrac{1}{2} g_{jk} b^j b^k$ emerge, with no corresponding term involving $\tfrac{1}{2} g_{jk}  u^j u^k$. 
%; this can be traced to the different weightings of the $g(u,u)$ and $g(b,b)$ terms in (\ref{eqaction}), one with mass $m$, the other with volume $\mu$. 
For magnetic field, equation (\ref{eqlieb}) can be reexpressed as a familiar form of the induction equation, 
\begin{align}
\upartial_t b^i & + u^j \nabla_j b^i  - b^j \nabla_j u^i  + (\nabla_j u^j ) b^i =0 , 
\label{eqmomentum2d} 
\end{align}
for $\eta=0$, and similarly for equations such as (\ref{eqlies}, \ref{eqlierho}). We have considered the compressible case, as it does allow a clear emergence of  the magnetic pressure term. For the incompressible case one drops the internal energy $e(\rho,s)$ and in its place uses a Lagrange multiplier to enforce that the flow map conserves volumes, $\phi_* \mu = \mu$. %, as detailed in \cite{GiVa18}. 
The theory can be developed further from Lagrangian to Hamiltonian mechanics, using a non-canonical Poisson bracket as discussed in \cite{Mo82,Mo98}, \cite{MoAnPe20} and reviewed in \cite{We18}, with extensions to relativistic MHD \citep{DAMoPe15} and two-fluid MHD \citep{LiMiMo16}.

%
%\JVcom{You were worried about the para above being seen as a token; I think it's fine.}

%
%\AG{I was wondering if this paragraph --- driven by referee 3's comments --- really fits here, or at the beginning of the section, the introduction or the discussion... But I'm not sure it would fit in those other places quite. I guess I don't want the referee to think we're grudgingly acknowledging this by putting this material here....}
%
\subsection{Helicity} \label{ssechelicity}

Ideal fluid mechanics, MHD and extensions involve the Lie transport of vector fields in a flow, and resulting conserved quantities are the helicities that give the linkages between the integral curves or field lines of the various vector fields \citep[see, e.g.][and references therein]{ArKh98,Fu08,LiMiMo16,MoDo19}.  Helicities result from Noether's theorem under the symmetry of relabelling Lagrangian coordinates \citep{PaMo96}, and \cite{EnPeTo16} show that helicity is the only integral invariant of a general vector field transported in volume-preserving flows. Extensions to helicity and linkage in relativistic MHD are given by \cite{YoKaYo14}, and related to the relabelling symmetry in \cite{KaYoFu14}.

In ideal (non-relativistic) MHD we have conservation of magnetic helicity given by the integral of the helicity 3-form $h_\textrm{M} = \alpha \wedge \dd\alpha$ over $\Mc$ with appropriate boundary conditions.  Here $\alpha$ is a 1-form potential for $\bflux$ so that $\bflux = \dd\alpha$ and, from (\ref{eqlieB}) and in a suitable gauge,
\begin{align}
 \upartial_t \alpha & + \lie_u \, \alpha = 0  . 
\label{eqvp1} 
\end{align}
For this gauge the helicity form is conserved on fluid parcels
\begin{equation}
\upartial_t (\alpha \wedge \bflux) + \lie_u (\alpha \wedge \bflux ) = 0,  
\end{equation}
and its integral over $\Mc$ is independent of gauge with suitable boundary conditions. Explicitly, if we set 
%\JVcom{Changed the subscript M, C and K to roman.}
%
\begin{equation}
H_\textrm{M} = \int_{\Mc} \alpha \wedge B, 
\end{equation}
then we have 
\begin{equation}
\frac{\dd H_\textrm{M}}{\dd t}  = - \int_{\Mc} \lie_{u} \,( \alpha \wedge B) =  - \int_{\Mc} \dd[ u \ip (\alpha \wedge B)] = - \int_{\upartial \Mc}   u \ip (\alpha \wedge B) = 0 
\end{equation}
for $u$ parallel to the boundary $\upartial\Mc$. See \cite{ArKh98}, \cite{MoDo19} for more discussion, for example of gauge invariance. 
If the magnetic helicity density $h_{\textrm{M}} = \alpha \wedge B$ vanishes everywhere, the resulting Pfaff integrability condition means that locally $\alpha$ defines a family of surfaces $\Phi = \text{const.}$, with $\alpha \propto \dd\Phi$, and a further \emph{Godbillon--Vey helicity} invariant may be defined, as discussed  by \cite{We18}, \cite{WePrAnHu19} and \cite{Ma20}.

For the cross helicity, the appropriate 3-form is $h_\textrm{C} = \nu \wedge \bflux$ (in traditional terms $\uv\cdotv\bv$) and is conserved in incompressible, constant density flow. To see this here, scale out $\rho$ and write the momentum and induction equations as 
\begin{align}
 \upartial_t \nu & + \lie_u \, \nu  = -  \dd \pi   +  \lie_{\bvec} \, \bflat , \\
 \upartial_t \bflux & + \lie_u \, \bflux = 0 , 
\end{align}
where $\pi$ absorbs terms such as the magnetic pressure. Then we have 
\begin{align}
\upartial_t (\nu \wedge \bflux) + \lie_u ( \nu\wedge \bflux)  & = - \dd \pi \wedge \bflux + (\lie_{\bvec}\,  \bflat) \wedge \bflux  \\
 & = - \dd( \pi \bflux) + \lie_{\bvec} ( \bflat \wedge \bflux)  \\
 & = - \dd( \pi \bflux) + \dd[ {\bvec} \ip ( \bflat \wedge \bflux) ], 
 \label{eqhcevol}
\end{align}
using $\dd B = 0$, $\lie_{\bvec} \bflux = 0$ (as $\lie_{\bvec} \bvec =0$, $\lie_{\bvec}\mu = 0$ and $\bvec \ip \mu = \bflux$), and Cartan's formula. Again this leads to a conservation law; if we set 
\begin{equation}
H_\textrm{C} = \int_{\Mc} \nu  \wedge B, 
\end{equation}
we have
\begin{equation}
\frac{\dd H_\textrm{C}}{\dd t} = - \int_{\upartial \Mc}  \left[ - u \ip ( \nu \wedge \bflux) - \pi \bvec \ip \mu + \bvec \ip ( \bflat\wedge \bflux)  \right] = 0 , 
\end{equation}
given that $u, \bvec \parallel \upartial\Mc$. 

In MHD the kinetic helicity, the  integral of $\uv \cdotv\omegav$ in everyday terms, or here the integral of $h_\textrm{K} = \nu \wedge \zeta$ with $\zeta = \dd \nu$, is not conserved, but in incompressible flow obeys
\begin{align}
\upartial_t (\nu \wedge \zeta) + \lie_u (\nu \wedge \zeta)
& = - \dd \pi \wedge \zeta + ( \lie_{\bvec} \,\bflat) \wedge \zeta + \nu \wedge (\lie_{\bvec} \, \dd \bflat) \\
&  \simeq 2 (\lie_{\bvec} \, \bflat) \wedge \zeta \simeq 2 \nu \wedge (\lie_{\bvec}\,  \dd \bflat ), 
\end{align}
using $\simeq$ to discard $\dd(\cdot)$ terms. The two versions follow from
\begin{align}
 \dd (\lie_{\bvec} \, \bflat \wedge \nu ) & = \dd(\lie_{\bvec} \, \bflat) \wedge \nu -(\lie_{\bvec} \, \bflat) \wedge \dd\nu \\
 & = (\lie_{\bvec} \, \dd \bflat) \wedge \nu -(\lie_{\bvec} \, \bflat) \wedge \dd\nu \\ 
 & = \nu \wedge  (\lie_{\bvec} \, \dd \bflat) -(\lie_{\bvec} \, \bflat) \wedge \zeta. 
\end{align}
%

%If this is then integrated over $\Mc$ and suitable boundary conditions (zero magnetic field on $\upartial \Mc$ for example) employed, then we have conservation of the cross helicity, namely the integral of $\nu\wedge \bflux$ over $\Mc$. 

\section{Pull backs and Alfv\'en waves}

\subsection{Transformation under a mapping} 

In this section we again consider incompressible fluid flow of constant density $\rho$; we rescale the pressure $p$ and magnetic field $\bflux$ so as to absorb this quantity, effectively setting $\rho=1$. We restate the governing equations as 
\begin{align}
 \upartial_t \nu & + \lie_u \, \nu + \dd\pi  = \lie_\bvec  \,\bflat   ,  
\label{eqmomentum3} \\
\upartial_t \bvec  & + \lie_u \, \bvec  = 0 , 
\label{eqlieB3} 
 \end{align}
with
\begin{align}
& \nu = u_{\flat}, \quad \beta = b_{\flat}, \\
 & \pi  = p -  \tfrac{1}{2} g(u, u) + g(\bvec,\bvec),  \\
&  \divv  u   = 0 , \quad \divv \bvec = 0 . 
 \end{align}
In incompressible flow, both field $\bvec$ and flux $\bflux$ are Lie dragged from their initial conditions $\bvec_0$ and $\bflux_0$  by the flow map $\phi$ which preserves volume:
\begin{equation}
\bvec = \phi_* \bvec_0 , \quad \bflux = \phi_* \bflux_0, \quad \phi_* \mu = \mu.
\label{eqbpush}
\end{equation}
%
%this being the Cauchy solution. 
We will focus primarily on $\bvec$ below rather than $\bflux$. 

We now decompose the flow map $\phi$ as a composition of volume-preserving maps \citep{MaSh01}, 
\begin{equation}
\phi = \xi \circ \phib . 
\label{eqcompose}
\end{equation}
We have in mind a situation where there are waves on a background flow, in which case $\phib$ gives the Lagrangian mapping for the background flow $\ub$ with 
\begin{equation}
\ub = \dot{\phib} \circ \phib^{-1} 
\label{equbdef}
\end{equation}
and the map $\xi$ is the fluctuating flow map moving particles from the background flow to their final positions, with the velocity field $w$ defined for $\xi$ by 
\begin{equation}
w = \dot{\xi} \circ \xi^{-1} . 
\label{eqwdef} 
\end{equation}
We stress that as far as the mathematical development is concerned $\phib$ and $\xi$ are simply two mappings, volume-preserving diffeomorphisms of $\Mc$, neither being required to have any further properties; in particular the map $\xi$ can be of finite amplitude. Note with this that the bar over $\phib$ is at present just a label --- it is not a \emph{mean} map here in that we do not need to define any averaging process %(and in fact the notion of averaging an ensemble of maps is problematic)
 --- we prefer the adjective \emph{background} for this reason, rather than \emph{mean}. We are considering a single fluctuating map $\xi$; this will be replaced by a family of maps $\xi^\realab$ labelled by $\realab$ in due course, but at present it is a single map. 
%\JVcom{I changed the realisation label from $\alpha$ to $\realab$ to avoid confusion with the magnetic potential; I've defined a symbol so this can be changed to something else.} 
%\AG{Great idea - very happy with this!!} 
We note that in the literature the decomposition (\ref{eqcompose}) is sometimes referred to as the use of \emph{hybrid Eulerian--Lagrangian} (HEL) coordinates \citep[e.g.][]{RoSo06a,SoRo14}, in that the background fluid flow $\ub$ (from the map $\phib$) captures the Eulerian motion, while the family of maps $\xi^\realab$ captures subsequent Lagrangian displacements, which we refer to as \emph{fluctuations} for brevity.  

We can apply the pull back $\xi^*$ to carry fluid parcels from their positions given by $\phi$ at time $t$, to their positions in the background flow, given by $\phib$. This carries magnetic quantities according to the Cauchy solution and likewise for forms such as $B$ and $\nu$. For magnetic field we can define
\begin{equation}
 \bvecb = \xi^* \bvec = \phib_* \bvec_0, \quad
 \bfluxb = \xi^* \bflux = \phib_* \bflux_0,
\label{eqpullbackb} 
\end{equation}
where the second equalities, stemming from (\ref{eqbpush}), indicate that the pulled back fields $\bvecb$ and $\bfluxb$ are 
just the push forwards of the initial condition by the background flow map $\phib$ and so are independent of $\xi$. For the flow,  differentiating (\ref{eqcompose}) shows that 
\begin{equation}
u = \xi_* \ub  + w , \quad \ub = \xi^* ( u - w) . 
\label{eqpullbacku} 
\end{equation}
Thus the background flow $\ub$, pushed forward by $\xi$, and the fluctuation velocity $w$ sum to give the flow velocity at any point. 

%Just to clarify our notation: we use a bar to indicate quantities that depend only on the mean map $\phib$ (and maybe some initial condition); these are independent of the choice of fluctuating map $\xi$. For quantities that depend on $\xi$ we will use a tilde. 

For other quantities we set
\begin{equation}
\nuh = \xi^* \nu, \quad 
\pih  = \xi^* \pi, \quad
\bflath = \xi^* \bflat , \quad
\gh = \xi^* g, \quad
\muh = \xi^* \mu , 
\end{equation}
so that the tilde is a label for these and other quantities obtained by applying $\xi^*$. Note that the background fields $\ub$ and  $\bvecb$ depend only on the background flow map $\phib$  (and for $\bvecb$ the initial condition), but this is not the case for $\nuh$, $\pih$, $\bflath$, hence the distinct notation. This becomes relevant below when we have an ensemble of maps, say $\xi^\realab$.

Applying $\xi^*$ to the equations for $\nu$ and $b$ and using standard identities \citep{GiVa18}  gives 
\begin{align}
 \upartial_t \nuh & + \lie_{\ub} \nuh + \dd\pih  = \lie_{\bvecb}  \,\bflath   ,  
\label{eqmomentum4} \\
\upartial_t \bvecb  & + \lie_{\ub} \bvecb  = 0 . 
\label{eqlieB4} 
 \end{align}
The induction equation just simplifies to (\ref{eqlieB4}), namely motion of the background field $\bvecb$ in the background flow $\ub$, in other words all trace of the fluctuations has vanished. However for the momentum equation, originally in (\ref{eqmomentum3})  we had transport of $\nu = u_{\flat}$ in the flow $u$:  momentum $\nu$ and velocity $u$ were simply related by the metric $g$ at each point. Now, in (\ref{eqmomentum4}) the momentum and flow are not so easily related. We have instead that
\begin{equation}
\nuh = \xi^* \nu = \xi^* [g(u, \cdot)] =  \xi^* [g(\xi_* \ub + w , \cdot)] =  \gh (\ub, \cdot) + \gh ( \wh, \cdot), 
\label{eqnuh}
\end{equation}
and for $\beta$, 
\begin{equation}
\bflath = \xi^* \bflat = \xi^*[ g(\bvec, \cdot)]  = \gh (\bvecb, \cdot) , 
\label{eqbflath}
\end{equation}
and so the relation involves the background flow for $\nuh$ and the pulled-back metric $\gh = \xi^* g$ for both. Thus, the momentum equation involves transport, not of the momentum of the background flow which would be $(\ub)_\flat$ but transport of a quantity $\nuh$ that differs from this, and likewise for $(\bvecb)_\flat$ and $\bflath$. For the magnetic field, the difference $\gh (\bvecb, \cdot)  - (\bvecb)_{\flat}$ between the two expressions captures the difference between turning the vector field $\bvecb$ into a 1-form field using the metric $g$ or the pulled back metric $\gh$.
%
%In the magnetic case the two terms give the change in the geometry from the fluctuating map, namely $\gh (\bvecb, \cdot)  - (\bvecb)_{\flat}$, that is the effect of going from vector field $\bvecb$ to 1-form field with the actual metric $g$ versus the pulled back metric $\gh$. 
For the momentum $\nuh$, there is an additional term arising from the fluctuating flow itself. 

\subsection{Lagrangian averaging} 

In this section we will make some comments about how the above framework is used for Lagrangian averaging, with an emphasis on MHD; see \cite{Ho02a} and \cite{GiVa18} for further discussion. We suppose that we have an ensemble of fluctuating maps $\xi^\realab$, labelled by $\realab$, on the same background flow given by $\phib$ or $\ub$. We also assume that we have the same initial magnetic field in each realisation of the ensemble, so that barred quantities such as $\ub$, $\bfluxb$ and $\bvecb$ are independent of $\realab$. On the other hand, tilded quantities do depend on $\realab$ and we can label these with an $\realab$ to stress this if we need to (we do this sparingly). For any tensor quantity, say $\tau^\realab$. we can define its Lagrangian average as
\begin{equation}
\barL{\tau} = \langle \xi^{\realab *} \tau^\realab \rangle, 
\label{eqbarLdef}
\end{equation}
where  $\langle\cdot\rangle$ denotes an ensemble average. For barred quantities, we have
\begin{equation}
\barL{\bvec} = \bvecb \quad \textrm{and} \quad  
\barL{\bflux} = \bfluxb, 
\end{equation}
by virtue of (\ref{eqpullbackb}). In contrast, 
\begin{equation}
\barL{u} \not= \ub,
\end{equation}
in general, since $\ub$ is defined in terms of the map $\phib$ by (\ref{equbdef}) and not in terms of a Lagrangian average;  as argued by \cite{GiVa18}, the equality cannot hold for any useful definition of the  mean flow map $\bar \phi$.
Note that in (\ref{eqbarLdef}) any tensor quantity is transported from the locations $\phi^\realab (x)$ of a single Lagrangian parcel in each realisation to a single point $\phib(x)$ in $\Mc$ and then averaged. Thus averaging always takes place in the tangent space, co-tangent space and tensor spaces at each point. We never attempt to average vectors and tensors located at different points of $\Mc$, as this cannot be defined in general. (The generalised Lagrangian mean theory of \cite{AnMc78a,AnMc78b} (see also \cite{Bu09})  assumes a  Euclidean structure to average vectors based at different points using parallel translation.)

 Applying this Lagrangian average to the incompressible MHD equations (\ref{eqmomentum4}, \ref{eqlieB4}), and letting barred quantities come out of any averaging, we have 
\begin{align}
 \upartial_t \barL{\nu} & + \lie_{\ub} \barL{\nu} + \dd\barL{\pi}  = \lie_{\bvecb}  \,\barL{\bflat}   ,  
\label{eqmomentum5} \\
\upartial_t \bvecb  & + \lie_{\ub} \bvecb  = 0 . 
\label{eqlieB5} 
 \end{align}
The Lagrangian averaged momentum $\barL{\nu}$ and $\barL{\beta}$ differ from the similar background quantities  by what are known as the \emph{pseudomomentum} $\pseudop$ and the \emph{pseudofield} $\pseudoh$ respectively; these are given by 
%%
%\begin{equation}
%-\pseudop =  \nuh - (\ub)_\flat = \gh (\ub, \cdot)  - (\ub)_{\flat} + \gh ( \wh, \cdot) , \quad
%-\pseudoh = \bflath - (\bvecb)_{\flat}  =  \gh (\bvecb, \cdot)  - (\bvecb)_{\flat}
%\end{equation}
%%
%(the minus sign for $\pseudop$ is a convention; see \cite{AnMc78a}). In the magnetic case the two terms give the change in the geometry from the fluctuating map, namely $\gh (\bvecb, \cdot)  - (\bvecb)_{\flat}$, that is the effect of going from vector field $\bvecb$ to 1-form field with the actual metric $g$ versus the pulled back metric $\gh$. For the pseudomomentum, there is an additional term from the fluctuating flow itself. In the context of GLM theory, \cite{Ho02a} also identifies $\pseudoh$ and refers to its average, over an ensemble of flows, as the \emph{magnetization induced by Lagrangian averaging}. 
%
\begin{equation}
-\pseudop =  \barL{\nu} - (\ub)_\flat , \quad
-\pseudoh = \barL{\bflat} - (\bvecb)_{\flat} . 
\end{equation}
(the minus sign for $\pseudop$ is a convention; see \cite{AnMc78a}). In the context of GLM theory, \cite{Ho02a} also identifies $\pseudoh$ and refers to its average, over an ensemble of flows, as the \emph{magnetization induced by Lagrangian averaging}. The pseudomomentum and pseudofield capture the effect of the fluctuations on the background flow; explicit expressions in terms of $\xi^\realab$ can be obtained by averaging (\ref{eqnuh}, \ref{eqbflath}). We give an example in the next section. 

We finally consider the effect of a pull back and Lagrangian averaging on the various helicities discussed in section  \ref{ssechelicity}. We consider first cross helicity \citep{Ho02a}, applying a pull back to equation (\ref{eqhcevol}) to obtain
\begin{align}
\upartial_t (\nuh \wedge \bfluxb) + \lie_{\ub} ( \nuh\wedge \bfluxb)  & =  - \dd( \pih \bfluxb) + \dd[ {\bvecb} \ip ( \bflath \wedge \bfluxb) ]; 
\end{align}
taking an average gives 
\begin{align}
\upartial_t (\barL{\nu} \wedge \bfluxb) + \lie_{\ub} ( \barL{\nu}\wedge \bfluxb)  & =  -\dd( \barL{\pi} \bfluxb) + \dd[ {\bvecb} \ip (\barL{\bflat} \wedge \bfluxb) ]. 
\end{align}
Thus we can define the Lagrangian averaged cross helicity form by $\barL{h_\textrm{C}} = \barL{\nu}\wedge\bfluxb$ and note that it differs from that of the background fields, namely $(\ub)_\flat \wedge \bfluxb$ by a term involving the pseudomomentum, namely $- \pseudop \wedge \bfluxb$. The cross helicity form obeys the above transport equation. This means that for any subvolume $\Vc$ of $\Mc$ we can account for the change of cross helicity in $\Vc$ by fluxes across the boundary $\Sc = \upartial \Vc$
\begin{equation}
\upartial_t \int_{\Vc} \barL{\nu}\wedge\bfluxb = \int_{\Sc} \left[ - \ub \ip (\barL{\nu}\wedge \bfluxb)  - \barL{\pi} \bfluxb + {\bvecb} \ip (\barL{\bflat} \wedge \bfluxb ) \right] .
\end{equation}

Applying a Lagrangian average to the magnetic helicity simply gives  $\barL{h_\textrm{M}}={\alphab} \wedge \dd {\alphab}$, since the  potential $\alpha$ is transported by the flow: $\xi^*\alpha = {\alphab} = \phib_* \alpha_0$, so nothing is gained by Lagrangian averaging: no linkages are changed in ideal evolution.
%so we do not gain anything over what we know from the initial conditions with the vector potential $\alpha$ transported in our present gauge as $\xi^*\alpha = {\alphab} = \phib_* \alpha_0$. 
Applying a Lagrangian average to the kinetic helicity gives $\barL{h_\textrm{K}} = \barL{\nu \wedge \zeta}$, which does not seem open to any useful simplifications in MHD, as $\zeta=d\nu$ is no longer given by the Cauchy solution. In pure hydrodynamics ($\bflux=0$), if all realisations in the ensemble have the same initial vorticity field (maybe only locally, if not globally) we can write $\xi^* \zeta = \zetab = \phib_* \zeta_0$ and so $\barL{h_\textrm{K}} = \barL{\nu} \wedge {\zetab}$. 
Again this differs from the kinetic helicity of the background fields, namely $(\ub)_\flat \wedge \zetab$, by a term involving the pseudomomentum, $- \pseudop \wedge \zetab$, Helicity conservation is expressed by
\begin{equation}
\upartial_t \int_{\Vc} \barL{\nu}\wedge\zetab = \int_{\Sc} \left[ - \ub \ip (\barL{\nu}\wedge \zetab)  - \barL{\pi} \zetab \, \right] , 
\end{equation}
which is zero if the background vorticity $\zetab$ vanishes on $\Sc$. Thus, for cross helicity in MHD and for kinetic helicity in hydrodynamics, the helicity `hidden' in the fluctuations is easily expressed in terms of the pseudomomentum. 

\subsection{Alfv\'en waves} \label{ssecalfvenwave} 

As a fundamental example we consider how a pull back can be employed for an Alfv\'en wave on a uniform magnetic field in Euclidean geometry, $\Mc = \Rbb^3$ with coordinates $(x,y,z) = (x^1, x^2, x^3)$ and the usual metric $ g = \dd x^2 + \dd y^2 + \dd z^2$. In ideal MHD with a uniform background magnetic field, waves of arbitrary shape can propagate without change, with the velocity and disturbance magnetic field proportional to each other, as found by \cite{Wa44} \citep[see, e.g.][]{AlFa50}. We consider a flow and magnetic field giving such a travelling wave by setting
\begin{equation}
f = f(z-ct), \quad h = h(z-ct) , 
\end{equation}
and then 
\begin{align}
u  & =  f\,\upartial_x  + h \,\upartial_y + \U \,\upartial_z, \quad \bvec  = \pm  f\,\upartial_x  \pm h \,\upartial_y + \B  \,\upartial_z, \label{eqalfvenflow}\\
\nu  & =  f\,\dd x + h \,\dd y + \U \,\dd z,  \quad \bflat  = \pm  f\,\dd x  \pm h \,\dd y + \B \, \dd z, 
\end{align}
where $\U$ and $\B$ are constant uniform flow and field respectively, $f$ and $h$ are arbitrary functions, and the wave speed
\begin{equation}
c = \U \mp \B. 
\label{eqcvals} 
\end{equation}
%
%\JVcom{I changed the font of $U$ and $B$ to avoid the clash of the fixed scalar $B$ with the 2-form $B$. I thought it'd be nice to show some calculation details to illustrate the powers of forms (remove if you don't like my missionary zeal).}
It is easy to check that this satisfies (\ref{eqmomentum3}, \ref{eqlieB3}) with $\pi = 0$ by manipulating the 1-forms $\nu$ and $\beta$ directly, using the commutation of Lie and exterior derivatives:
%It is easily checked that this satisfies (\ref{eqmomentum3})--(\ref{eqlieB3}) with $\pi = 0$; Cartan's formula (\ref{eqCartan}) is helpful here. 
\begin{align}
\upartial_t \nu + \lie_u \nu &= -c (f' \, \dd x + h' \, \dd y ) + \lie_u f \,  \dd x + \lie_u h \,  \dd y + \lie_u \U \,  \dd z + f \, \dd (\lie_u x) + h \, \dd(\lie_u y) + \U \, \dd (\lie_u z) 
\nonumber \\
&= (\U-c) f' \, \dd x + (\U-c) h' \, \dd y + f \, \dd f + h \, \dd h,  \\
\lie_b \beta &= \pm \lie_b f \, \dd x  \pm \lie_b h \,\dd y + \lie_b \B \,  \dd z \pm f \, \dd (\lie_b x) \pm h \, \dd (\lie_b y) + \B\,  \dd (\lie_b z) \nonumber\\
&= \pm \B f'\,  \dd x \pm \B h' \, \dd y + f \, \dd f + h\,  \dd h  . 
\end{align}
%
%\JVcom{I rearranged a bit the section below: I thought the logic was a bit confusing: the form of $F$ and $H$ is settled once $\bar u$ is chosen, so I think it's better to give it right away.}
We now map the vector field $u$ in (\ref{eqalfvenflow}) onto a background flow chosen as $\ub = \U \,\upartial_z$; in other words, we remove the Alfv\'en wave by the pull back. This is achieved with fluctuations of the form 
\begin{equation}
\xi (x,y,z,t) = ( x + F(z-ct), y + H(z-ct), z). 
\end{equation}
and associated  velocity 
\begin{equation}
w = -c F' \, \upartial_x - c H' \,  \upartial_y. 
\end{equation}
We can use (\ref{eqpullbacku}), that is, $\bar u = \xi^* (u - w)$, to compute $\bar u$. Noting that
\begin{equation}
\pxh = \xi^* \upartial_x  = \upartial_x , \quad
\pyh = \xi^* \upartial_y = \upartial_y, \quad
\pzh = \xi^* \upartial_z = - F' \, \upartial_x - H' \upartial_y + \upartial_z,
\end{equation}
we find
\begin{equation}
\ub   = (  f \mp \B F') \, \upartial_x +(  h \mp \B H') \, \upartial_y  + \U \, \upartial_z ,
\end{equation}
hence we choose
\begin{equation}
F' = \pm \B^{-1} f, \quad   H' = \pm \B^{-1} h. 
\end{equation}
With this choice, the background magnetic field, obtained from (\ref{eqpullbackb}) as
\begin{equation}
\bvecb   = ( \pm f - \B F') \, \upartial_x +( \pm h - \B H') \, \upartial_y  + \B \, \upartial_z,
\end{equation}
also simplifies, leading to the uniform background vector fields
\begin{align}
\ub  &  =  \U \, \upartial_z , \quad 
\bvecb  = \B \, \upartial_z , 
\end{align}
which trivially satisfy the induction equation (\ref{eqlieB4}).
The corresponding 1-forms, in contrast, are more complicated. Using that
\begin{equation}
\dxh = \xi^* \dd x  = \dd(\xi^* x) = \dd x + F'\, \dd z, \ \
\dyh =  \xi^* \dd y  = \dd(\xi^* y) = \dd y + H'\, \dd z, \ \ 
\dzh =  \xi^* \dd z   = \dd(\xi^* z) = \dd z,   
\label{eqpullbackd}
\end{equation}
we obtain
\begin{align}
\nuh  &  =  f\, \dd x + h\, \dd y +  [\U \pm \B^{-1} (f^2 + h^2)] \, \dd z,  \label{eqnuh1} \\
\bflath &  = \pm f \, \dd x \pm h \, \dd y + [ \B + \B^{-1} (f^2 + h^2)] \, \dd z,  \label{eqbflath1}
\end{align}
which depend explicitly on the waves through $f$ and $h$. It can be checked that (\ref{eqnuh1}, \ref{eqbflath1}) solve the pulled-back momentum equation (\ref{eqmomentum4}).

If we have a sea of Alfv\'en waves with random phases such that $\langle f \rangle = \langle h \rangle  = 0 $, then we can take a Lagrangian average  as outlined above to replace previously tilded quantities, and obtain 
\begin{align}
\barL{\nu} &  =  [\U \pm \B^{-1} \langle f^2 + h^2 \rangle]\, \dd z,   \quad
\barL{\bflat}   = [  \B + \B^{-1} \langle f^2 + h^2\rangle]\, \dd z, \\
 -\pseudop  &  =    \pm \B^{-1} \langle f^2 + h^2\rangle \, \dd z,   \quad
-\pseudoh   =      \B^{-1}  \langle f^2 + h^2 \rangle \, \dd z  . 
\end{align}
Thus we gain a correction to the $z$-directed momentum of the background flow $(\ub)_\flat = \U\, \dd z$ and to the mean 1-form field $(\bvecb)_\flat = \B \, \dd z$ that parameterises these waves, and we identify the pseudomomentum and pseudofield as quantities that are quadratic in the wave amplitude, and have only a $z$ component.  The cross helicity  
carried by the waves (with the background contribution subtracted out) is
\begin{equation}
- \pseudop \wedge \bfluxb =  \pm  \langle f^2 + h^2\rangle \,\dd x\wedge \dd y \wedge \dd z. 
\end{equation}
The above example is straightforward since the Alfv\'en waves are wholly transverse and the nonlinear terms cancel out; the results are equivalent to what would be obtained by GLM theory and averaging over the phases of the waves. Nonetheless this example does capture how a mapping can straighten out magnetic field lines and encode the disturbance in the pseudo-quantities. Extension to non-uniform magnetic fields and compressible flows, in order to understand wave--mean flow/field interactions \citep[see][]{Bu09}, are the subject of future study, beyond the scope of the present paper.

%Discussion of helicity, cross helicity and energy 

\section{The Braginsky dynamo}

As a second example of the use of a geometric approach to MHD, we consider the Braginsky dynamo. We will allow for compressible flow $u$ in this section. As a kinematic dynamo, the flow field $u$ is specified and for non-zero, uniform magnetic diffusion $\eta>0$, the induction equation (\ref{eqlieB}) becomes
\begin{equation}
\upartial_t \bflux  + \lie_u \, \bflux    = \eta \nabla^2 \bflux.
\label{eqlieB6}
\end{equation}
Here, the operator $\nabla^2$ is  (minus) the Laplace--de Rham operator,  
\begin{equation}
- \nabla^2 = \delta \dd + \dd \delta,
\end{equation}
where $ \delta \equiv \star \dd \star  $ is the \emph{codifferential operator}, here mapping the 2-form $B$ to a 1-form.
We have already identified $j = \dd \bflat = \dd {\star}\bflux$ as the current 2-form, and thus the operator $\dd \star$ corresponds to a curl. Since $\dd\bflux=0$, 
\begin{equation}
\nabla^2  \bflux = - \dd \delta  \bflux = - \dd {\star} \dd {\star}  \bflux
\end{equation}
is equivalent to the two curls in (\ref{eqeverydayind}, \ref{eqeverydaymisc}). 
%
%where   . Note that the appropriate viscous diffusion operator is different in a general setting, as it arises from the divergence of a stress tensor; see the discussion in \cite {GiRiTh14}, \cite{GiVa20a}. 
% JV: I think this is better omitted.

%\AG{Yes - checking Frankel around p 369 we have that in a 3-d manifold ${\star}{\star} \alpha = \alpha$ for all $p$-forms $\alpha$, that in 3-d the codifferential acting on 2-forms is $ \delta = d^* = {\star} d {\star}$ (there is a minus sign for 1-forms!), and $\Delta = - \nabla^2 = d d^* + d^*d$. }

In kinematic dynamo theory, the aim is to show that growing solutions exist for a particular choice of flow and diffusivity $\eta>0$. %We give a sketch for the reader who has some familiarity  with $\alpha\omega$-dynamos, aiming to work in the most general geometry and, in so doing, to isolate just those aspects that are suited to a geometric treatment. 
Our goal is to obtain formulae for the transport term known as the $\alpha$-effect, by exploiting the machinery of pull-backs and operators in a general setting, rather than the explicit construction of $\alpha\omega$-dynamos, or explicit calculation of $\alpha$ for particular families of  fluctuations or waves. We stress that the overarching ideas behind the $\alpha$-effect and $\alpha\omega$-dynamos are well established and it is only our approach that is novel; for comprehensive discussion, as well as solutions in particular examples, we refer to the book \cite{MoDo19}.

\subsection{Informal approach: an $\alpha\omega$ dynamo} \label{ssecmotivation}

We first illustrate the ideas underlying the Braginsky dynamo and their geometric formulation by considering a specific scenario. We use coordinates $(x^1, x^2 , x^3) = (x,y,z)$; these need not be Cartesian, nor need $(x,y)$ even be orthogonal, but we do assume that the metric is independent of $z$ taking the form
\begin{equation}
g(x,y)  = \begin{pmatrix} \;\; g_{11} \;& \; g_{12} \; & 0 \\ g_{21} & g_{22} & 0 \\  0 & 0 & \;\; g_{33}\;\;  \end{pmatrix} .
\label{eqgsimple} 
\end{equation}
Thus $g$ is invariant under translations in the $z$-direction and under the transformation $z \to - z$; examples include spherical polar coordinates $(r, \theta, \phi)$ with $z\equiv 
\phi$ the azimuthal angle or longitude. The volume form $\mu = |g|^{1/2} \, \dd x \wedge \dd y \wedge \dd z$ is also $z$-independent. For convenience we set $\varsigma(x,y)  = |g|^{-{1/2}}$, with $|g|$ the determinant of $g$, so that the induced volume form is   
\begin{equation}
\mu = \varsigma^{-1}  \, \dd x\wedge \dd y \wedge \dd z. 
\end{equation}
We let the magnetic  flux 2-form $\bflux$ be written as 
\begin{equation}
\bflux = \bflux_1 \,\dd y \wedge \dd z  +  \bflux_2 \, \dd z \wedge \dd x  + \bflux_3 \, \dd x \wedge dy. 
\label{eqbfluxdef}
\end{equation}
Note that when we use index notation below we will also use $B_{ij}$ with $B_1 = B_{23} = - B_{32}$ etc., that is $B = \tfrac{1}{2} B_{ij} \, \dd x^i \wedge \dd x^j$, there being little risk of confusion.  
%\JVcom{Added a factor 1/2 and used $\wedge$; not sure what you meant by `though apologies are owed to the summation convention' -- maybe that a factor 1/2 was missing.}. 
The condition that $B$ is solenoidal, $\dd B=0$, amounts simply to
\begin{equation}
\upartial_x B_1 + \upartial_y B_2 + \upartial_z B_3 =0 , 
\end{equation}
valid for any coordinates. 
We need the Hodge star operator $\star$, which in three dimensions gives an isomorphism between $1$-forms and $2$-forms involving the metric and volume form. Starting with the basis 2-forms, we readily find  
\begin{equation}
\dd y \wedge \dd z \overset{\star}{\longleftrightarrow} g_{11} \varsigma \, \dd x + g_{21} \varsigma\, \dd y, \quad
\dd z \wedge \dd x  \overset{\star}{\longleftrightarrow} g_{12} \varsigma \, \dd x + g_{22} \varsigma\, \dd y, \quad
\dd x \wedge \dd y  \overset{\star}{\longleftrightarrow} g_{33} \varsigma \, \dd z. 
\label{eqexplicithodge} 
\end{equation}
We are now ready to discuss the operation of an $\alpha\omega$-dynamo, which we build step by step.

First consider a flow in the $z$-direction only, of the form 
\begin{equation}
u = u^3(x,y)\, \upartial_z . 
\label{equsimple} 
\end{equation}
The induction equation (\ref{eqlieB6}) with $\eta=0$ becomes
% \JVcom{Again, I thought a bit of explicit computation with forms would be useful.}
%
\begin{align}
0&=(\upartial_t + \lie_u) (\bflux_1 \, \dd y \wedge \dd z  +  \bflux_2 \, \dd z \wedge \dd x  + \bflux_3 \, \dd x \wedge \dd y) \nonumber \\
&=[(\upartial_t + u^3 \upartial_z) \bflux_1 ]\, \dd y \wedge \dd z + \bflux_1 \, \dd y \wedge \dd u^3 + [(\upartial_t + u^3 \upartial_z) \bflux_2 ]\, \dd z \wedge \dd x + \bflux_2 \, \dd u^3 \wedge d x \nonumber \\
&  +[ (\upartial_t + u^3 \upartial_z) \bflux_3] \, \dd x \wedge \dd y,
\end{align}
hence, in components,
\begin{equation}
(\upartial_t + u^3\upartial_z ) \bflux_1 = 0 , \quad 
(\upartial_t + u^3\upartial_z ) \bflux_2 = 0 , \quad 
(\upartial_t + u^3\upartial_z ) \bflux_3 =  (\bflux_1 \upartial_x + \bflux_2 \upartial_y ) u^3 . 
\label{equsimpleb} 
\end{equation}
On the left-hand side we have transport of field components in the flow $u$, and on the right-hand side of the final equation we observe the well-known $\omega$-effect, namely the generation of the azimuthal $\bflux_3$ component from transverse components $\bflux_1$ and $\bflux_2$. Given the simplicity of the flow field, there is no term that regenerates $\bflux_1$ and $\bflux_2 $ field. 

We now include diffusion $\eta>0$ and see what diffusive processes can convert $\bflux_3(x,y)$ field into $\bflux_1$, $\bflux_2$ components. Making use of (\ref{eqexplicithodge}), we calculate the right-hand side of (\ref{eqlieB5}) applied only to this component as follows:
\begin{align}
\bflux & = \bflux_3(x,y) \, \dd x\wedge \dd y, \\
\star \bflux   & =   g_{33} \varsigma \bflux_3\, \dd z , \\
\dd {\star} \bflux   & =    (g_{33} \varsigma \bflux_3)_y \,\dd y \wedge  \dd z  -   (g_{33} \varsigma \bflux_3)_x \,\dd z \wedge  \dd x , \\
\delta \bflux = {\star} \dd {\star} \bflux  
& = [ g_{11} \varsigma  (g_{33} \varsigma \bflux_3)_y -   g_{12} \varsigma  (g_{33} \varsigma \bflux_3)_x ]\,  \dd x 
+ [ g_{21} \varsigma  (g_{33} \varsigma \bflux_3)_y -   g_{22} \varsigma  (g_{33} \varsigma \bflux_3)_x ] \, \dd y , \\
 \dd \delta \bflux = \dd {\star} \dd {\star} \bflux
& =   
% - [ g_{21} \varsigma  (g_{33} \varsigma \bflux_3)_x -   g_{22} \varsigma  (g_{33} \varsigma \bflux_3)_y ]_z  \dd y\wedge \dd z \\
% & + [ g_{11} \varsigma  (g_{33} \varsigma \bflux_3)_x -   g_{12} \varsigma  (g_{33} \varsigma \bflux_3)_y ]_z \dd z \wedge \dd x \\
\bigl[ [ g_{21} \varsigma  (g_{33} \varsigma \bflux_3)_y -   g_{22} \varsigma  (g_{33} \varsigma \bflux_3)_x ]_x - [ g_{11} \varsigma  (g_{33} \varsigma \bflux_3)_y -   g_{12} \varsigma  (g_{33} \varsigma \bflux_3)_x ]_y  \bigr] \, \dd x\wedge \dd y . 
\label{eqD3calc} 
\end{align}
%
%Could give this for $g_{12} = g_{21}=0$ and comment that it works more generally. 
Thus, the diffusion of $\bflux_3(x,y)$ field in this geometry does not lead to $\bflux_1$, $\bflux_2$ components as defined in (\ref{eqbfluxdef}). 

If the $\bflux_3$ field depends additionally on $z$ then there is the generation of $\bflux_1$, $\bflux_2$ components, but a field depending on $z$ is liable to enhanced diffusion because of the effect of $u^3(x,y) $ in (\ref{equsimple}) in reducing transverse scales and enhancing dissipation. Instead, aiming for an $\alpha\omega$-dynamo in the traditional formulation, we specify a  field $\bflux_3(x,y)$ that is independent of $z$ and so robust to this process, and seek other mechanisms. In the Braginsky dynamo, the idea is to replace the flow in (\ref{equsimple}) by something more complicated, namely by adding some finite amplitude $z$-dependent motion to the flow field.

Within the context of our geometric approach, we consider now the flow map $\phi$ for the velocity field $u$ to be written as 
\begin{equation}
\phi = \xi \circ \phib, 
\label{eqphidecompose}
\end{equation}
where the background flow $\ub = \dot\phib \circ \phib^{-1}$ again takes the simple form 
\begin{equation}
\ub = \ub_3 (x,y) \, \upartial_z, 
\label{equbsimple}
\end{equation}
and $\xi$ is a general, time-dependent, finite amplitude map; in particular $\xi$ can have arbitrary $z$-dependence.  We can have in mind for example that the full flow $u$ consists of finite-amplitude waves on a simple background shear flow $\ub$. Applying $\xi^*$ to the induction equation (\ref{eqlieB6}) gives us the pulled back version 
\begin{equation}
\upartial_t \bfluxh  + \lie_{\ub}\,  \bfluxh  = - \eta\,  \xi^* (\dd \delta \bflux )  .
\label{eqlieBpb1b}
\end{equation}
We use $\bfluxh = \xi^* \bflux$ rather than $\bfluxb$ in an earlier section, since with diffusion $\eta>0$ we no longer have the Cauchy solution to relate the magnetic field to its initial condition. 
Equations (\ref{equsimpleb}) are now relevant for the effect of the flow $\ub$ (\ref{equbsimple}) on the field $\bfluxh$ in the absence of dissipation, $\eta=0$, and we have the $\omega$-effect acting on the $\bfluxh_1$ and $\bfluxh_2$ components to generate a $\bfluxh_3$ component; the effect of the distortions to the flow via $\xi$ has completely vanished from the left hand side. The effect though is present on the right-hand side as the Hodge star operator involves the volume form and metric (as in (\ref{eqexplicithodge})), and under the pull back we can write 
\begin{equation}
\xi^* (\dd \delta  \bflux) = \xi^* ( \dd {\star}\dd {\star}  \bflux )  = \dd \starh \dd \starh   \bfluxh , 
\label{eqxiddelta} 
\end{equation}
where $\starh$ applies the star using the pulled back metric $\gh$ and volume form $\muh$. We recall how the Hodge star operator works, here in this pulled-back version. We take a 2-form field $\bfluxh$ and generate the corresponding vector field $\bvech$ via $\bvech \ip \muh = \bfluxh$. We then use the flat operator to give us a 1-form field $\bflath$ via $\bflath = \gh ( \bvech, \cdot)$. This field is $\bflath = \starh \bfluxh$ and provided the pulled back metric $\gh$ is sufficiently complicated, i.e.\ the fluctuations coded in $\xi$ break enough symmetry, then we would expect 
(\ref{eqxiddelta}) in (\ref{eqlieBpb1b}) to generate $\bfluxh_1$ and $\bfluxh_2$ components  from $\bfluxh_3(x,y)$ -- the origin of the $\alpha$-effect. 

To make further progress we write these Hodge star calculations in coordinates as:
\begin{equation}
\bvech{}^i \, \varsigmah^{-1}\,  \eps_{ijk} = \bfluxh_{jk}, \quad 
\bvech{}^i = \varsigmah \,  \tfrac{1}{2}\, \eps^{ijk}\,  \bfluxh_{jk} , \quad 
\bflath_i = \gh_{ij}\,  \bvech{}^j, \quad
\bflath{}_i = \gh_{ij} \,   \varsigmah \, \tfrac{1}{2} \,\eps^{jkl} \,\bfluxh_{kl}, 
\label{eqexplicithodgestar}
\end{equation}
%
%or 
%%
%\begin{equation}
%\bvech{}^i \ =  \tfrac{1}{2} \varsigmah \eps^{ijk} \bfluxh_{jk}, \quad 
%\bflath_i = \gh_{ij} \bvech{}^j, \quad
%\bflath{}_i = \varsigmah \gh_{ij}  \bfluxh_{j}.
%\end{equation}
%%
where $\eps_{ijk}$ or $\eps^{ijk}$ is the usual Levi--Civita alternating symbol, and we recall the discussion below (\ref{eqbfluxdef}) about components written as $\bfluxh_i$ versus $\bfluxh_{jk}$. Also we have for the action of $\dd$ on any 1-form $\lambda$,
\begin{equation}
(\dd \lambda )_{ij} = 2 \upartial_{[i} \lambda_{j]}  = \upartial_i \lambda - \upartial_j \lambda, 
\label{eqdcomp}
\end{equation}
where $[\cdot]$ is antisymmetrisation. With this we can write (for any magnetic flux 2-form $\bfluxh$) the components of the (pulled back) \emph{electromotive force}%
\footnote{This should perhaps be referred to as an \emph{effective} emf as it emerges from the current as a component of the magnetic diffusion term, with $ \emf = - \eta \, {\star} j$, $\emfh = - \eta\, {\starh} \tilde{\jmath}$, and so includes all the usual diffusive effects, rather than being defined from taking the average of $\uv \times\bv$ in the traditional formulation; see \cite{SoRo14}.}
(emf) $\emfh$, defined by 
\begin{equation}
\emfh = - \eta \, \xi^*(\delta  B) = - \eta \,\starh \dd \starh  \bfluxh,
\end{equation}
as
\begin{align}
\emfh_i  %- \eta \,(\starh d \starh \, \bfluxh)_{i} 
& =-  \eta \,\gh_{ij} \,\varsigmah \, \eps^{jkl} \,\upartial_{[k}(\gh_{l]m} \,\varsigmah\, \tfrac{1}{2} \,\eps^{mnp} \, \bfluxh_{np}) \\
& =  -  \eta\, \gh_{ij} \,\varsigmah\, \eps^{jkl} \,\upartial_{k}(\gh_{lm}\, \varsigmah )  \,\tfrac{1}{2} \,\eps^{mnp} \, \bfluxh_{np}  -  \eta \, \gh_{ij}\, \varsigmah\, \eps^{jkl} \, \varsigmah\, \gh_{lm}\,\tfrac{1}{2} \eps^{mnp} \,\upartial_{k} \bfluxh_{np} 
\label{eqfulldiff} 
\end{align}
(in the last line the antisymmetrisation is dropped, being redundant because of the $\eps^{jkl}$ term).  So far this is exact, providing the diffusion operator for any distortion $\xi$ of the coordinate system taking the background flow $\ub$ and field $\bfluxh$ to the actual, wavey fields, $u$ and $\bflux$. All the complexity is of course hidden in the pull back $\xi^*$, giving the tilde fields, as it involves the coordinate map and its derivatives. 

We now consider as input a field $\bfluxh = \bfluxh_3 (x,y) \, \dd z$, with $\bfluxh_{12} =  - \bfluxh_{21} = \bfluxh_3$, etc. This corresponds to taking $m=3$ in the above equation (and $n$, $p$ are $1$ or $2$). We are also interested in sources of field for the  transverse $\bfluxh_1$ and $\bfluxh_2$ components, which corresponds to taking $i=3$ in the above, as we shall see. Thus, our focus is on:
\begin{align}
\emfh_3 = - \eta\, (\starh d \starh \, \bfluxh)_{3} 
&  = -  \eta\,\gh_{3j} \,\varsigmah \,\eps^{jkl} \,\upartial_{k}(\gh_{l3} \,\varsigmah )  \, \bfluxh_{3}  -  \eta \,\gh_{3j} \,\varsigmah \,\eps^{jkl}\,  \varsigmah \,\gh_{l3} \, \upartial_{k} \bfluxh_{3} 
\label{eqdstar6}\\
&  =  - \eta\, \gh_{3j} \,\varsigmah \,\eps^{jkl} \,\upartial_{k}(\gh_{l3} \,\varsigmah )  \, \bfluxh_{3} .
\label{eqdstar7}
\end{align}
The second term on the right hand side of (\ref{eqdstar6}) vanishes by symmetry to leave (\ref{eqdstar7}); the remaining  term in (\ref{eqdstar7})  vanishes if the metric $\gh$ takes the unperturbed form (\ref{eqgsimple}), as indeed it must, but in the presence of some non-trivial distortion of the coordinate system $\xi$ %, i.e.\  a wave on the background flow, 
will generally be non-zero. Suppose finally that we have a family of such waves, given by $\xi^{\realab}$, that are translation invariant in $z$ (while $\bfluxh_3$ is independent of $z$ as above). Then if we average the term over such waves we obtain a quantity in (\ref{eqdstar6}, \ref{eqdstar7}) that is $z$-independent and takes the form 
\begin{align}
\emfh_3 = - \eta \, \langle (\starh \dd \starh  \,   \bfluxh )_{3} \rangle &  =   \alpha    \bfluxh_{3}   ,  
\end{align}
with 
\begin{align}
\alpha (x,y) =-  \eta\,  \langle\,\gh_{3j}\, \varsigmah \,\eps^{jkl} \,\upartial_{k}(\gh_{l3} \,\varsigmah ) \rangle .
\label{eqalpha3}
\end{align}
Since
\begin{equation}
\dd (\emfh_3 \, \dd z) = \dd ( \alpha \bfluxh_3 \, \dd z) =  (\alpha \bfluxh_3)_y \, \dd y \wedge \dd z -  ( \alpha \bfluxh_3)_x \, \dd z\wedge \dd x, 
\end{equation}
this $\alpha$-effect term gives the required coupling from the $\bfluxh_3$ field to the $\bfluxh_1$ and $\bfluxh_2$ components (and note that the other components $i=1$, $2$ of  $\emfh_i = - \eta \, \langle \starh \dd \starh \, \bfluxh \rangle_{i}$ would not, since the average yields a quantity independent of $z$). 

If we take all the magnetic field components to be independent of $z$ we gain the governing equations as
\begin{align}
&  \upartial_t \bfluxh_1 =    (\alpha \bfluxh_3)_y   - \eta  D_1,  
\label{eqalom1}
\\
& \upartial_t \bfluxh_2 = -  ( \alpha \bfluxh_3)_x  - \eta D_2 ,  
\label{eqalom2}
\\
& \upartial_t  \bfluxh_3 =  (\bfluxh_1 \upartial_x + \bfluxh_2 \upartial_y ) u^3  -  \eta D_3 . 
\label{eqalom3}
\end{align}
where the terms $D_i$ are the remaining diffusion terms, as discussed further below.
%we gain, as per the calculations in (\ref{eqD3calc}) and equivalent (we omit the details which are straightforward). 
Several remarks are in order in relating our results to the usual theory of kinematic dynamos. First we have extracted the $\alpha$-effect generating $\bfluxh_1 $ and $\bfluxh_2$ components from the $\bfluxh_3$ component of the magnetic flux: these $\alpha$-effect terms are crucial because they give the feedback loop that enables an $\alpha\omega$-dynamo to function. We have not sought coupling terms that for example generate a $\bfluxh_3$ component from a $\bfluxh_1$ or $\bfluxh_2$ component, but this is not really necessary as we are assuming an $\omega$-effect, the terms $(\bfluxh_1 \upartial_x + \bfluxh_2 \upartial_y ) u^3$ in the $\bfluxh_3$ equation (\ref{eqalom3}), to be present already. Typically in an $\alpha\omega$-dynamo the $\alpha$-effect is relatively weak, giving components $\bfluxh_1, \bfluxh_2 \ll \bfluxh_3$ and so any additional $\alpha$ type couplings would be a subdominant effect; if this is not the case then one can develop the theory of $\alpha^2$- or $\alpha^2\omega$-dynamos, in which case these couplings need to be quantified. 

%We outline a more general case below, but here have focussed on the simplest setup. 

%Without these extra complexities, 
%the curl of the emf (here given by $d\mathcal{E}$) feeds back into the induction equation, which could be written as 
%%
%\begin{equation}
%\upartial_t \bfluxh  + \lie_{\ub} \bfluxh  =  d (\alpha \bfluxh_3\,  \dd z) - \eta\,  d {\star} d {\star} \bfluxh  .
%\label{eqlieBpb1a}
%\end{equation}
%%
%with, as we indicated above, replacement of the full diffusion operator by the easier term $ - \eta\,  d {\star} d {\star} \bfluxh$ in many applications. 

% In the general geometric framework as here, the possibilities proliferate and we will not investigate this further. 

The key point of this approach, as recognised by earlier authors going back to \cite{So72}, is that applying a pull back keeps the structure of the advection and stretching terms  in (\ref{eqlieBpb1b}) (those not involving $\eta$) and it is not necessary to average these: all the averaging is done in the diffusion term. Expanding the diffusion term allows  an $\alpha$-effect to emerge that is not present in the original equations and that parameterises how fluctuations superposed on a background flow diffusively generate other magnetic flux components. 
%Of course in the remaining diffusion terms $D_i$  there will be many more terms that arise from the fluctuations: however these will give corrections to effects that are already present. For example there will be a correction to the calculation in (\ref{eqD3calc}). However we often care little, if at all, about this for two reasons: first that it does not make any difference to the physical effect --- diffusion with a correction is still pretty much diffusion --- and second that in most frameworks in which one might calculate the $\alpha$-effect, these corrections would be small and so negligible at leading order. However we do care about the $\alpha$-effect terms identified in (\ref{eqalom1}, \ref{eqalom2}) as   these neatly parameterise the effect of fluctuations and the $z$ dependence of the true field $\bflux = \xi_* \bfluxh$, to regenerate transverse field diffusively and so close the dynamo loop. 
Of course the identification of an $\alpha$-effect was a key contribution to dynamo theory, in this context in the seminal papers of  \cite{Br64a,Br64b} with parallel work by Parker, Steenbeck, Krause and R\"adler as reviewed in \cite{MoDo19}. In short, introducing the above $\alpha$-effect and ignoring other effects of the fluctuating flow leads to the classic induction equation augmented by the $\alpha$ term (\ref{eqalpha3}), as (\ref{eqalom1}--\ref{eqalom3}) or 
\begin{equation}
\upartial_t \bfluxh  + \lie_{\ub}\,  \bfluxh  =  \dd (\alpha \bfluxh_3\,  \dd z) - \eta\,  \dd {\star} \dd {\star} \bfluxh  .
\label{eqlieBpb1a}
\end{equation}

\subsection{General case}\label{ssecdiscussalpha}

Let us now return to the general setting of the induction equation 
\begin{equation}
\upartial_t \bflux  + \lie_u \,  \bflux  = - \eta\, \dd \delta \bflux =  \dd \emf,  \quad \emf =- \eta\, \delta \bflux ,   
\label{eqlieB7}
\end{equation}
with any metric $g$ and flow $u$, where $\emf $ is again the 1-form emf.  We apply the earlier decomposition $\phi = \xi \circ \phib$ in (\ref{eqphidecompose}) and apply a pull back $\xi^*$ to the equation so as to remove some fluctuating component of the flow --- as usual we have in mind waves on a simpler background flow $\ub$ generated by $\phib$. The resulting equation is as in (\ref{eqlieBpb1b}):
\begin{equation}
\upartial_t \bfluxh  + \lie_{\ub}\,  \bfluxh  =  \dd \emfh  . 
\label{eqlieBpb1c}
\end{equation}
Here as above $\emfh$ is the pulled-back emf given by 
\begin{equation}
 \emfh =   \xi^* \emf =  - \eta\, \xi^* ( \delta \bflux) = - \eta\, \xi^* [\delta (\xi_* \bfluxh)] =  - \eta \,   \tilde{\delta} \bfluxh  ,    
\label{eqlieBpb1cemf}
\end{equation}
where we recall that $\bfluxh = \xi^* \bflux$ is the pulled back field and where the second equality here defines the pullback $\tilde{\delta} = \xi^*\delta\xi_* = \starh \dd \starh$ of the codifferential operator $\delta$. 

We now follow this by averaging over a family $\xi^{\realab}$ of fluctuations to obtain 
\begin{equation}
\upartial_t  \barL{\bflux} + \lie_{\ub}\,  \barL{\bflux}   =  \dd\barL{\emf } , \quad \barL{\emf } = - \eta\, \langle  \tilde{\delta} \bfluxh \rangle = - \eta \, \barL{\delta \bflux}, 
\label{eqlieBpb1c1}
\end{equation}
where
\begin{equation}
\barL{\bflux} = \langle \bfluxh \rangle = \langle \xi^* B \rangle.  
\end{equation}
There is no approximation up to this point.
% \JVcom{I feel we need to define $\barL{\bflux}$; the previous appearance is many pages ago and in a form specific to the zero-conductivity case, when no averaging is necessary.} 
This expression for the mean emf $\barL{\emf}$ involves both the magnetic field $\bfluxh$ and the fluctuations through $\deltah$ 
and in an exact development the product $\barL{\delta \bflux}$  generally cannot be broken up into the product of two averages.  We therefore write  $\bfluxh$ as the sum of its (Lagrangian) mean and fluctuating component, setting
\begin{equation}
\bfluxh = \barL{\bflux} + \bfluxh^{\ell} , \quad \bflux = \xi_* \barL{\bflux} + \bflux^\ell, \quad \bfluxh^{\ell} = \xi^* \bflux^{\ell}, 
\end{equation}
and writing 
\begin{equation}
\barL{\emf } = - \eta\,\barL{\delta} \barL{B} - \eta \, \barL{\delta \bflux^{\ell}} .
\label{eqemfsplit} 
\end{equation}
Of these two terms, the first, $ - \eta\,\barL{\delta} \barL{B} $, represents a transport effect: the effect of the operator $\barL{\delta}$ on a mean field $\barL{B}$, as discussed informally in the previous section. Our aim is to obtain expressions for this operator, which includes an $\alpha$-effect and effective diffusivity. We will not consider the second term, $ \eta \, \barL{\delta \bflux^{\ell}}$ as it would usually be handled separately in a dynamo calculation and be neglected at leading order. As in any transport problem, one calculates the coefficients or operators governing the transport of a mean field, and this is then fed into a more global calculation which involves further assumptions, such as scale separation, or a parameter being large or small, in order to neglect some terms. In the case of the Braginsky dynamo, a limit of large magnetic Reynolds number, here $\eta\ll1$, results in the fluctuating fields $\bfluxh^{\ell}$ being negligible, even for large Lagrangian displacements encoded in $\xi$ and so in $\barL{\delta}$. %\JVcom{Is this really true? I don't see why small conductivity leads to small perturbations.}

For these reasons we will neglect the second term in (\ref{eqemfsplit}) and focus on the first, our aim being to elucidate the structure of the averaged operator $\barL{\delta}$. To avoid introducing new quantities we will use $\doteq$ to signal that the second term in (\ref{eqemfsplit}) has been dropped, and  write 
\begin{equation}
\barL{\emf }  \doteq 
- \eta\,\barL{\delta} \barL{B}.
\end{equation}
The operator $\barL{\delta} = \langle \tilde \delta \rangle$ involves derivatives and we will expand this shortly as 
\begin{equation}
\barL{\emf} \doteq - \eta \, \barL{\delta} \barL{\bflux} =  \alpha \ip \barL{\bflux} + \gamma \ip \nabla \barL{\bflux} 
\end{equation}
or, in coordinates, 
%\JVcom{The coordinate-free expression above is pretty and is better here than in the perturbative section where you had it.}
%
\begin{equation}
\barL{\emf }_i  \doteq 
%(- \eta\,\barL{\delta} \barL{B} )_i  = 
\alpha_i^{np} \, \barL{B}_{np} + \gamma_i^{knp}\,  \nabla_k  \barL{B}_{np}.
\label{eqemfbarLB}
\end{equation}
Here the $\alpha_i^{np}$ are the components of the $\alpha$ tensor, and the $\gamma_i^{knp}$ are the components of the effective difusivity tensor.  Our approach below will be to work first with the pull back $\emfh = \deltah \bfluxh$ and then ensemble average: at that point we can replace $\bfluxh$ with $\barL{\bflux}$ and extract the $\alpha$-effect and $\gamma$ tensor.

We now derive an explicit expression for $\emfh$ in a way that sidesteps much of the coordinate computations of section \ref{ssecmotivation}.
This relies on two observations. First, we use the relation between the codifferential $\delta$ and the covariant derivative $\nabla$ of differential forms \citep[][Theorem 14.15]{Fr97} which, when applied to the 2-form $B$, gives
\begin{equation}
(\delta B)_i = g^{jk}\,  \nabla_k B_{ij}.
\label{eqdeltanabla}
\end{equation}
For completeness we establish this property in Appendix \ref{appdivergence}. It makes it possible to rewrite $\emfh$ in (\ref{eqlieBpb1cemf}) as
\begin{equation}
\emfh_i      =- \eta\,  \gh^{jk}\,  \nablah_k \bfluxh_{ij} ,
\label{eqemf3}
\end{equation}
where we have introduced the pull back $\nablah = \xi^*\nabla$ of the covariant derivative $\nabla$. This is defined in a coordinate-free way by 
\begin{equation}
\nablah_u \tau \equiv (\xi^* \nabla)_u \tau = \xi^* \left[ \nabla_{\xi_* u} (\xi_* \tau )\right] 
\end{equation}
\citep{St17}. In other words on the right-hand side we push $u$ and $\tau$ forwards, apply $\nabla$ and then pull the result back. This gives $\nablah$ as the covariant derivative with respect to the pulled back metric $\gh$. Rewriting (\ref{eqemf3}) as
\begin{equation}
\emfh_i    =  - \eta\,     \gh^{jk}\,  (\nablah_k - \nabla_k) \bfluxh_{ij}  - \eta\,    \gh^{jk}\,  \nabla_k \bfluxh_{ij} , 
\label{eqfullalpha8a}
\end{equation}
we make the second observation: the difference of covariant derivatives, applied to any tensor, just  returns  a tensor (no derivatives are involved) involving connection coefficients $C^{i}_{jk}$ (symmetric in $j$ and $k$) with, for example, 
\begin{equation}
(\nablah_k - \nabla_k ) v^j = C^j_{lk} \, v^l, \quad
(\nablah_k - \nabla_k ) \lambda_j = -  C^l_{jk} \, \lambda_l, 
\label{eqconnection}
\end{equation}
for  a vector field $v$ and a 1-form field $\sigma$  \citep{St17}. 
%\AG{Note that I took the opposite sign of $C$ to that in Stein (15) as it then looks better for the Christoffel symbol version....}
Here the coefficients $C^{i}_{jk}$ are the difference between the Christoffel symbols for the two metrics $\gh$ and $g$, 
\begin{equation}
C^i_{jk} = \tilde{\Gamma}^i_{jk} -  \Gamma^i_{jk}. 
\end{equation}
Using these and noting that
\begin{equation}
(\nablah_{k} - \nabla_k)  \bfluxh_{ij} =-  C^s_{ki} \, \bfluxh_{sj} -   C^s_{kj} \, \bfluxh_{is} , 
\label{eqnablediffB}
\end{equation}
we write
\begin{align}
\emfh_i   & = \eta\,    \gh^{jk}\, ( C^s_{ki} \, \bfluxh_{sj}+   C^s_{kj} \, \bfluxh_{is} -  \tilde g^{pk} \nabla_k \bfluxh_{ip})  .
% \\
 % & =   \alpha_i^{np} \bfluxh_{np}  +  \gamma_i^{knp} \nabla_{k} \bfluxh_{np}. 
  \label{eqalphagamma}
\end{align}
Bearing in mind (\ref{eqemfbarLB}) and the discussion there, we now replace $\bfluxh$ by the mean field $\barL{\bflux}$ and ensemble average, meaning we only average over the metric $\gh$ and the connection coefficients $C^i_{jk}$ and not the field, to identify the tensors $\alpha$ and $\gamma$ as
\begin{align}
\alpha_i ^{np} & =  \eta \, \langle  C^{[n}_{ki}\, \gh^{p]k} + \gh^{jk}\,  \delta_i^{[n} \, C^{p]}_{jk} \rangle 
\label{eqfullalpha8b} \\
 \gamma_i^{knp} &= - \eta \,\delta_i^{[n} \, \langle \gh^{p]k} \rangle \label{eqgammagamma1}
\end{align}
(the $[np]$ antisymmetrisation is optional as the $np$ indices are to be contracted against the 2-form magnetic field).  
%\JVcom{Needs more on the meaning of $\alpha$ and $\gamma$ and sorting out the averaging.}
This gives an equation for the $\alpha$ tensor as derived from the tensor $C^{i}_{jk}$ giving the difference between the connection coefficients of the covariant derivative and the pulled back covariant derivative. This fundamental geometric formulation of $\alpha$ is new as far as we are aware. 

%OLD STUFF 

%We stress that the formulae for the general tensor components $\alpha_i^{np}$ in (\ref{eqfullalpha8b}), and likewise for the $\alpha$ component in (\ref{eqalpha3}) above and the more classical version in (\ref{eqfullalpha3a}), are exact and cover finite amplitude fluctuations. 
%They include compressible or incompressible flow, %(for the latter, $\varsigmah = \varsigma$ is independent of the fluctuations), 
%this being the principal benefit of working with the magnetic flux 2-form $\bflux$ rather than the vector field $\bvec$ here. In appendix  \ref{appbragex} we confirm that this gives the correct $\alpha$-effect for the case of a simple helical wave, and in appendix \ref{appexplicitalpha} we relate our approach to some well-established formulae in the literature, for example given in \cite{MoDo19}.}

%\JVcom{The above para is a bit disordered, citing (C.2) before mentioning Appendix C. I propose:

% NEW STUFF
We stress that  formula (\ref{eqfullalpha8b})  for the general tensor components $\alpha_i^{np}$ is exact and covers finite amplitude fluctuations. It includes compressible or incompressible flow, %(for the latter, $\varsigmah = \varsigma$ is independent of the fluctuations), 
this being the principal benefit of working with the magnetic flux 2-form $\bflux$ rather than the vector field $\bvec$ here. In appendix  \ref{appbragex} we confirm that (\ref{eqfullalpha8b}) gives the correct $\alpha$-effect for the case of a simple helical wave. 
In appendix \ref{appexplicitalpha} we relate our approach to some well-established formulae in the literature, for example given in \cite{MoDo19}. We start by stating an alternative, more classical expression for  $\alpha_i^{np}$, given as (\ref{eqfullalpha3a}), which holds with the same generality as (\ref{eqfullalpha8b}). % but is restricted to three dimensions. 
We verify the equivalence of this expression with (\ref{eqfullalpha8b}) and use it to recover the Euclidean results of \cite{RoSo06a,RoSo06b}.

%\JVcom{Should we emphasise that (\ref{eqfullalpha8b}) is new, and not simply a generalisation to arbitrary coordinates of the `classical' formula (\ref{eqfullalpha3a})  which generalises (\ref{eqalpha3}). I feel this would help deal with some of the referees comments about novelty. In my mind, it's more or less clear that (\ref{eqfullalpha8b}) is more fundamental and geometric than (\ref{eqfullalpha3a}), which for instance holds only in 3D.}

\subsection{Perturbative calculation of the $\alpha$-effect}
%
%\AG{This section is following your comment about the pull back of the Hodge star operator and the Trautman reference. It actually chimes with stuff I did thinking about this a few years ago, but didn't have the machinery that we now have. Of course this is really standard in the usual Cartesian coordinates, etc., but I haven't seen the differential geometry set-up.}
%
%\JVcom{I think the two sections on perturbation remain a bit problematic: it still a lot of computations to take in for the reader without a clear narrative. Perhaps they should be merged in two subsections, with each described as the perturbative counterpart of the two formulas we have for $\alpha$, one obtained using $\delta$, the other using $\nabla$.} 

All the above discussion is for finite amplitude maps $\xi$ and requires no small parameter in order to write down formulae for the $\alpha$-effect and effective diffusivity. In a realistic application the waves or fluctuations encoded  in $\xi$ would be developed in powers of a small parameter $\eps\ll1$ as far as needed to obtain the leading non-zero $\alpha$-effect, together with the limit of small diffusivity, here $\eta\ll1$, in the full dynamo calculation \citep{SoRo14}. We can take two approaches to calculate $\alpha$ perturbatively for small amplitude $\eps\ll1$. The first (\S4.3.1) is to take the pulled-back codifferential $\deltah = {\starh} \dd {\starh}$, and to expand the pulled-back Hodge star operator $\starh = {\xi^*} {\star}$ as a power series in the amplitude of the fluctuations, this involving the Lie derivative of $\star$ in a vector field $q$ used to generate the fluctuation map $\xi$ at any time. The second approach (\S4.3.2) works from (\ref{eqfullalpha8a}), which employs a pulled-back covariant derivative $\nablah$ in the difference $(\nablah_k-\nabla_k)\bfluxh_{ij}$. This difference of derivatives, which relates to  the conection coefficient tensor $C^i_{jk}$, can likewise be expanded as a power series in the amplitude of the fluctuating map $\xi$.  
%These two approaches give different but equivalent formulae for the $\alpha$-effect. 
%The equivalence of these formulae to the more traditional expressions in the literature is set out in appendix C.
%\JVcom{The last sentence is inaccurate: there's nothing in appendix C about the perturbative results. I propose: 
These two approaches give different but equivalent formulae for the $\alpha$-effect corresponding to perturbative expansions of (\ref{eqfullalpha3a}) and (\ref{eqfullalpha8b}), respectively.

\subsubsection{Perturbative calculation based on $\delta$} \label{ssecpertalpha}

 Let us return to the more abstract setting, and write the emf as 
\begin{equation}
\barL{\emf} \doteq - \eta \, \barL{\delta} \barL{\bflux} =   - \eta \, \langle \deltah \rangle \barL{\bflux} = - \eta\,  \langle \starh \dd \starh \rangle \barL{\bflux} 
= \alpha \ip \barL{\bflux} + \gamma \ip \nabla \barL{\bflux}. 
\label{eqTraut1}
\end{equation}
Here the derivatives in $\langle\delta\rangle$ and the $\langle{\star}\dd{\star}\rangle$ act on quantities to their right, including the mean magnetic field $\barL{B}$. We can develop $\xi(x)$ as a perturbation series, writing
\begin{equation}
\xi^i(x)  = x^i + \eps q^i(x) + \tfrac{1}{2} \eps^2 q^j(x)\, \upartial_j q^i(x)  + \cdots 
\end{equation}
(we suppress the dependence of all these quantities on time $t$). 
%We will only go to terms of order $\eps$ in this series, but note that if such series are taken to quadratic order or beyond, it is helpful to think of 
Here $q$ is a vector field defined on $\Mc$ (generally depending on time) which, when integrated over a fictitious time variable $s$ from `time' $s=0$ to $s=\eps$ (at any fixed time $t$), effects the map $\xi$ \citep{SoRo10,GiVa18}, or formally $\xi = \exp ( \eps q	)$. For simplicity we take $q$ to depend on $(x,t)$ but not on $s$: it is steady in fictitious time. At leading order a pull back is then given for any tensor $\tau$ by 
\begin{equation}
\xi^* \tau = \tau +\eps \lie_q\,  \tau + \tfrac{1}{2} \eps^2  \lie_q \lie_q \, \tau + \cdots , 
\end{equation}
from the definition of the Lie derivative. Now suppose we expand (\ref{eqTraut1}) in powers of $\eps$ to obtain 
\begin{align}
 \barL{\emf}   \doteq - \eta\,  \langle \starh \dd \starh \rangle \barL{\bflux} & =   - \eta \, \bigl[ \star \dd {\star}  + \eps \langle\lie_q \star\rangle   \dd {\star}   +  \eps\, {\star}  \dd \langle\lie_q  \star  \rangle  
 \notag\\
 &   + 
\tfrac{1}{2} \eps^2 \langle\lie_q \lie_q \star\rangle   \dd {\star}   +  \eps^2 \langle\lie_q {\star}   \,\dd \, \lie_q {\star}  \rangle + \tfrac{1}{2}  \eps^2 \, {\star}  \dd \langle\lie_q \lie_q  {\star}  \rangle  +  \cdots \bigr] \barL{\bflux}, 
\label{eqTraut2}
\end{align}
where we can think of the star operator as simply the tensor with components $\star_i^{kl}  = \tfrac{1}{2} g_{ij} \,\mu^{jkl}$ giving the map from 2-forms to 1-forms in this context; see (\ref{eqexplicithodgestar}). %; see (\ref{emfagain1}), say. 

There is a proliferation of terms in (\ref{eqTraut2}); however in typical applications many will be zero by virtue of the symmetry of the underlying system. For example in our sketch $\alpha\omega$-dynamo we restricted to a metric with a simple structure (\ref{eqgsimple}), independent of the third coordinate, and extracted an $\alpha$-effect taking $B_3$ field to $\emf_3$, needed to close the dynamo loop. In a `working' dynamo model specific choices would need to be made, and if the fluctuations encoded in $q$ lead to averages such as $\langle \lie_q {\star}\rangle$ and $\langle \lie_q \lie_q {\star} \rangle$ retaining a simple structure then the terms involving these averages above may well be zero, or not contribute to the components of the $\alpha$ tensor needed in that dynamo. For example in the informal approach of section 4.1, the key term is (\ref{eqalpha3}) and involves an average with the exterior derivative $d$ sandwiched between two quantities. 
%, because of the antisymmetry induced by the action of $d$. 
%\JVcom{It's not clear why these terms are more likely to vanish than the one you retain. Can this be made explicit in the set of the `informal approach'?}
Bearing in mind that this would need to be checked on a case-by-case basis, we will %take such terms to vanish here and 
calculate just the key term involving $d$ similarly sandwiched inside  the average,  writing
\begin{align}
 \barL{\emf}  \doteq  - \eta\,   \eps^2 \langle\lie_q {\star}  \, \dd \, \lie_q {\star}  \rangle   \barL{\bflux} . 
 %= \alpha\ip \barL{\bflux} + \gamma \ip \nabla \barL{\bflux} .
\label{eqTraut2a}
\end{align}
To calculate $\lie_q{\star}$ we follow the development in \cite{Tr84}. Define the tensor $h$ by 
\begin{equation}
\lie_q \, g = h g 
\quad\text{or}\quad  
(\lie_q g)_{ij} = h_i^{\,k} \,g_{kj} .  
\label{eqlieqg1}
\end{equation}
%
%Then we have
%%
%\begin{equation}
%(\lie_q \mu)^{ijk} = - \tfrac{1}{2} (\trace h)  \,\mu^{ijk} , \quad \trace h = h_i^{\,i}, 
%\end{equation}
%%
%%\JVcom{The formula above is confusing because $\mu$ doesn't mean the volume form but its dual. What's true is
%\begin{equation}
%\lie_q \mu = + \tfrac{1}{2} (\trace h)  \,\mu. 
%\end{equation}
%%Can we write $\mu$ in your equation as $\tfrac{1}{6} \mu^{ijk} (d x^i)^\sharp \wedge (d x^j)^\sharp \wedge (d x^k)^\sharp$?
%}\AG{Yes --- will try to fix this.... point taken....}
%%
%%\AG{How about simply the following? Noting that I have used this tensor a lot from (\ref{emfagain1}) onwards, and it would be odd at this late point to put down a definition. Of course one could argue that it should have beren done earlier....., though it was defined in words in a way that is probably clear enough for any readers to get that far...} 
Then we have
\begin{equation}
\lie_q \, \mu^{ijk} = - \tfrac{1}{2} (\trace h)  \,\mu^{ijk} , \quad \trace h = h_i^{\,i}, 
\end{equation}
and bearing in mind that $\star_i^{kl}  = \tfrac{1}{2} g_{ij} \,\mu^{jkl}$,  it follows that 
\begin{equation}
\lie_q\,  {\star} = ( h - \tfrac{1}{2} \trace h)\, {\star}
\quad\text{or}\quad  
(\lie_q \, \star)_i^{jk} = ( h^{\,l}_i - \tfrac{1}{2}  \delta_i^l  \trace h) \, \tfrac{1}{2} g_{lm} \,\mu^{mjk} .
\end{equation}
From (\ref{eqlieqg1}), the tensor $h$ is plainly linked to the \emph{deformation tensor} $\sigma$, that is the rate of change (Lie derivative) of the metric in the flow $q$, with 
\begin{equation}
\sigma \equiv  h g= \lie_q\,  g  = \nabla q_\flat + (\nabla q_\flat)^{\mathrm{T}} 
\quad\text{or}\quad  
\sigma_{ij} \equiv h_i^{\,k} \,g_{kj}= (\lie_q \, g)_{ij} = \nabla_i q_j + \nabla_j q_i. 
\end{equation}
%
%\AG{actually maybe $\sigma$ is not the best letter to use here...} 
We also need the fact that in the definition of $d$ acting on a 1-form in (\ref{eqdcomp}) partial derivatives may be replaced by covariant derivatives (given that there is no torsion for a covariant derivative induced by a metric), and so we have for any 1-form $\lambda$
\begin{equation}
(\dd \lambda )_{ij} = 2 \nabla_{[i} \lambda_{j]}  = \nabla_i \lambda_j - \nabla_j \lambda_i . 
\label{eqdcompcov}
\end{equation}

With these definitions, the $\alpha$ tensor can be written in a variety of ways, 
\begin{align}
\alpha_i^{np}  & = - \eta \, \eps^2 \,g_{rj} \,\mu^{jkl} \,  g_{sm} \,\tfrac{1}{2} \,\mu^{mnp} \langle (h^{\,r}_i - \tfrac{1}{2}  \delta_i^r  \trace h)\, \nabla_{k} (h^{\,s}_l - \tfrac{1}{2}  \delta_l^s  \trace h) \rangle
\label{eqalphafinal1}
\\
 & = - \eta \, \eps^2\, \mu^{jkl}\,\tfrac{1}{2}  \,\mu^{mnp} \langle(\sigma_{ij} - \tfrac{1}{2}  \,g_{ij}  \trace \sigma )\,\nabla_{k} (\sigma_{lm} - \tfrac{1}{2}  \,g_{lm}  \trace \sigma) \rangle
 \label{eqalphafinal2}
\\
  & = - \eta \, \eps^2\, \mu^{jkl}\,\tfrac{1}{2}  \,\mu^{mnp} \langle(\nabla_i q_j  + \nabla_j q_i - g_{ij}  \divv q )\,\nabla_{k} (\nabla_l q_{m}  + \nabla_{m} q_l - g_{lm}  \divv q )\rangle, 
  \label{eqalphafinal3}
\end{align}
with $\trace \sigma = \sigma_{ij} \, g^{ij} = 2 \divv q  =2  \nabla_i q^i$. Thus the $\alpha$-effect is expressed in a general form either in terms of the deformation tensor $\sigma$ or the vector field $q$ (and the corresponding 1-form $q_\flat$) generating the fluctuations and so the family of maps $\xi$. Note that if all maps are volume-preserving and flows incompressible we have $\trace \sigma=0$ and then we can write $\alpha$ compactly as 
\begin{align}
\alpha_i^{np}  &  = - \eta \, \eps^2\, \mu^{jkl}\,\tfrac{1}{2}  \,\mu^{mnp} \langle\sigma_{ij}  \,\nabla_{k} \, \sigma_{lm}  \rangle , \qquad (\trace \sigma=0). 
\label{eqalphafinal4}
\end{align}
There do not seem to be many general simplifications beyond this point, except to note that the term $\mu^{jkl}\nabla_{k} \nabla_l q_{m}$  in (\ref{eqalphafinal3}) involves the Riemann tensor and so vanishes if $\Mc$ is flat. 

%In our basic example $\eps q^1=\eps q_1 = F$, $\eps q^2= \eps q_2 =H$ and there is agreement with the previous formulae for the $\alpha$ tensor.

\subsubsection{Perturbative calculation based on $\nabla$}\label{ssecgenpert}

An alternative approach is to apply techniques in \cite{St17} to write down forms of the $\alpha$ tensor based on (\ref{eqfullalpha8a}), 
namely,
%which we repeat here, but dropping the tilde on the magnetic field for convenience: 
%
\begin{align}
\alpha_i ^{np}\, \barL{\bflux}_{np} & = - \eta \,\langle   \gh^{jk}  (\nablah_k - \nabla_k) \rangle \,\barL{\bflux}_{ij} . 
\label{eqfullalpha8c}
\end{align}
Since the averaged field $\barL{B}$ acts here as just a test field on which to do calculations, we replace it in this section only by $B$, to lighten notation:
%Although we have placed $\barL{\bflux}$ inside the average for convenience here and below, it is already a mean quantity and so can be taken outside too. 
%
\begin{align}
\alpha_i ^{np}\, B_{np} & = - \eta \,\langle   \gh^{jk}  (\nablah_k - \nabla_k) \rangle \,B_{ij} . 
\label{eqfullalpha8c1}
\end{align}

We introduce a series expansion for the $\alpha$ tensor, 
\begin{equation}
\alpha_i^{np} = \eps \alpha_{(1)i}^{np} +  \eps^2 \alpha_{(2)i}^{np} + \cdots , 
\end{equation}
with 
\begin{align}
&\alpha_{(m)i} ^{np} \bflux_{np} = - \frac{\eta}{m!}\,  \frac{\dd^m}{\dd s^m} \bigg|_{s=0}\, \Bigl\langle   (  \xi^* g^{jk})  \,  \xi^* ( \nabla_k \xi_* \bflux_{ij} )  - (\xi^* g^{jk})  \nabla_k  \bflux_{ij}   \Bigr\rangle  .
\label{eqalphalie}
\end{align}
Here, as above we suppose that the map $\xi$ is effected by a vector field $q$ over an interval of fictitious time $0 \leq s \leq \eps$; $q$ generally depends on time but not on $s$. We have for any tensor $\tau$, 
\begin{equation}
\lie_q \, \xi^* \tau = \frac{\dd}{\dd s} \, \xi^* \tau  , \quad 
 \lie_{-q} \, \xi_* \tau =  \frac{\dd}{\dd s} \, \xi_* \tau  . 
\end{equation}
Using these with (\ref{eqalphalie}) for $m=1$ and $m=2$, and averaging with $\langle \cdot\rangle$  gives after some algebra, 
\begin{align}
\alpha_{(1)i} ^{np} \bflux_{np} & =  - \eta \, g^{jk} \, [ \langle \lie_q \rangle , \nabla_k ] \bflux_{ij}  , \label{eqalphapert1}\\
\alpha_{(2)i} ^{np} \bflux_{np} & = -  \eta\,   \langle  ( \lie_q  \, g^{jk})  \,  [ \lie_q ,  \nabla_k] \rangle  \bflux_{ij}  - \eta  \, \tfrac{1}{2} g^{jk} \langle  [ \lie_q , [ \lie_q, \nabla_k ] ]  \rangle 
\bflux_{ij} , 
 \label{eqalphapert2}
\end{align}
where the square brackets denote a commutator, for example $[ \lie_q , \nabla_k ] \bflux_{ij} = \lie_q \nabla_k \bflux_{ij} -   \nabla_k \lie_q\bflux_{ij} $. All the terms in (\ref{eqTraut2}) are included here though, as mentioned above, in typical applications one would have $\alpha_{(1)i} ^{np} = 0 $ and many terms vanishing from $\alpha_{(2)i} ^{np}$. 
%\JVcom{I think the next sentence is very cryptic and would be better omitted.}
%Note that we have taken the flow $q$ independent of fictitious time $s$ though strictly we should allow $q$ this dependence so as not to restrict the type of diffeomorphism $\xi$ \citep[cf.\ discussion in][\S4.6]{GiVa18}; to do so gives a minor modification to $\alpha_{(2)i} ^{np}$ by bringing in new terms involving $q' = dq/ds$ at $s=0$. {\color{blue}(We remark that this doesn't affect the calculations in \S\ref{ssecpertalpha}.)}

\subsection{Summary}

We summarise the key results of section 4: we recall that these are stated using the magnetic field as a 2-form, and that the tilde annotation denotes any pulled-back quantity, that is transported from the flow with fluctuations or waves, to the background flow. The first, exact result is (\ref{eqfullalpha8a}, \ref{eqfullalpha8b}), which express the $\alpha$-effect 
\begin{align}
\alpha_i ^{np}\, \barL{\bflux}_{np} & 
= - \eta\, \langle   \gh^{jk}\,  (\nablah_k - \nabla_k)\rangle \barL{\bflux}_{ij}  
  = \eta\, \langle   \gh^{jk}\,  C^s_{ki} \rangle \, \barL{\bflux}_{sj} +  \eta \, \langle   \gh^{jk}\,  C^s_{kj} \rangle\, \barL{\bflux}_{is} , 
  %\\
% & =  \eta \, \langle  C^{n}_{ki}\, \gh^{pk} + \gh^{jk}\,  \delta_i^{n} \, C^{p}_{jk} \rangle  \, \mu_{npq}\, \bvech^q ,
 \end{align}
in terms of the difference between the covariant derivatives $\nablah$ and $\nabla$, or in terms of the connection coefficients $C^i_{jk}$, the difference in the connections between the metric $g$ and the pulled-back metric $\gh$.  
%In the final line we have expressed the effect in terms of  the magnetic vector field $\bvech$. 
%\JVcom{Could add: An alternative expression for $\alpha_i^{np}$, more closely related to earlier literature but restricted to three-dimensions, is given in (\ref{eqfullalpha8b}).}
This formula can be applied in space of any number of dimensions; an alternative expression for $\alpha_i^{np}$, more closely related to earlier literature but restricted to three dimensions, is given in (\ref{eqfullalpha3a}).

We give a variety of approximate formulae for the $\alpha$ tensor in sections \ref{ssecpertalpha} and \ref{ssecgenpert}  using a small-amplitude expansion in the magnitude $\eps$ of the fluctuations. In particular (\ref{eqalphafinal2}), %which may be written as %in terms of  $\bvech$ as 
\begin{align}
\alpha_i^{np} \, \barL{\bflux}_{np} 
& = 
 - \eta \, \eps^2\, \mu^{jkl}\,\tfrac{1}{2}  \,\mu^{mnp}  \langle(\sigma_{ij} - \tfrac{1}{2}  \,g_{ij}  \trace \sigma )\,\nabla_{k} (\sigma_{lm} - \tfrac{1}{2}  \,g_{lm}  \trace \sigma) \rangle\, \barL{\bflux}_{np}, 
 \label{eqalphalast}
\end{align}
expresses the $\alpha$-effect in terms of the deformation tensor $\sigma = \lie_{q} \, g$, that is the Lie derivative of the metric in the flow $q$ generating the fluctuations.

Note that we have preferred to work with the closed magnetic  2-form $B$ with $\dd B=0$ rather than the solenoidal magnetic vector field $b$ with $\divv b =0$, noting that $b\ip \mu = B$ and $\dd(b\ip\mu) = \mu \divv b$. For the general case of compressible flow $\dd B=0$ translates into $\dd\bfluxh=0$ and so a solenoidal mean field $\dd\barL{B}=0$. However in terms of $b$ we only obtain that $\dd (\barL{b\ip \mu})=0$ and so generally  $\divv \barL{b} \neq0$, in other words $\dd(\barL{b}\ip \mu) \neq0$: the Lagrangian average of $b$ is not solenoidal. This could be cured by recognising that $b$ transforms as a tensor of weight $-1$, but that is equivalent to using $B$ in any case. If all flows and maps are volume preserving, $\mu = \muh = \barL{\mu}$, this issue evaporates, $\divv \barL{b}=0$ and in the calculations above we can easily express the $\alpha$-effect in terms of the mean magnetic vector field $\barL{b}$ by replacing
\begin{equation}
\tfrac{1}{2}  \,\mu^{mnp} \, \barL{B}_{np} = (\barL{b}){}^m,
\end{equation}
in equations such as (\ref{eqalphalast}) above.

\section{Concluding remarks} 

In this paper, we have looked at MHD from a geometric perspective, both to derive the governing equations and their properties, and to revisit some basic applications, to Alfv\'en waves on a uniform field and to the analysis of the Braginsky dynamo. There are many attractions of a geometric approach. In concrete terms, results are valid for any coordinate system, and so no changes need to be made when going from, say, Cartesian coordinates to cylindrical polar coordinates, whereas otherwise this requires special consideration \citep{SoRo14}. Furthermore results remain correct even when one might find it helpful to adopt a non-orthogonal coordinate system, for example to use a buoyancy or pressure coordinate instead of a vertical coordinate in a geophysical or astrophysical setting. This flexibility is convenient even when working in $\Rbb^3$; in fact there is no difference in the theoretical development up to the point where second derivatives come in and curvature plays a role through the Riemann tensor, for example in a diffusion operator \citep{GiRiTh14}. For theoretical developments, the machinery of pull backs and push forwards allows one to apply mappings to equations while preserving their structure as much as possible. Essentially, a neighbourhood of any point in the interior of $\Mc$ is much like any other point from the viewpoint of the basic operations $\lie$, $\dd$, $\ip$ of differential geometry: we can take a calculation performed at one point and move it to another, provided we take the fields and all the necessary extra structure, that is the metric $g$ and volume form $\mu$, with us. This fact makes it possible to pin down why certain finite-amplitude approximations work, even if in real applications perturbation theory may well be needed for concrete calculations. 

There are some disadvantages; for example we have defined $\bflux$, $\bvec$ and $\bflat$, which are all the magnetic field in one version or another! To move between these we use the metric and/or volume form, and this allows the careful tracking of how quantities transform under mappings, pull backs or push forwards, applied to equations written in the form (\ref{eqlieB}) and (\ref{eqmomentum2a}). Note that using the `general relativity' notation (\ref{eqmomentum2c}, \ref{eqmomentum2d}), while it introduces fewer quantities, hides the underlying differential geometric structure, and applying a pull back   risks becoming entangled in transformations of Christoffel symbols, something generally worth avoiding. 

Concerning Alfv\'en waves, we addressed only the most fundamental model, and we plan to look into compressible waves, and waves on non-uniform fields, to see how these can be parameterised within the present framework, in parallel with similar developments in the literature of geophysical fluid dynamics; see, for example, \cite{Bu09}. One important point to note, as discussed in more detail in \cite{GiVa18}, and which emerges in other studies such as \cite{SoRo10}, is that given a family of flows with waves, there is no imperative need to define what the `mean flow' is at the outset. Although we use a bar (e.g.\ $\bar{u}$, $\bar{b}$), we refer to such fields as \emph{background} fields. In the decomposition (\ref{eqcompose}) which underlies all the work using pull backs, or equivalently hybrid Euler--Lagrangian coordinates or GLM-type theories, the choice of the mean map $\phib$ and flow $\ub$ is open, free to choose depending on applications. In our development, no assumption has been made along the lines that some sort of average of the fluctuations is zero, and in fact such assumptions are not easy to deploy in a general setting. Several possible choices of a mean flow \emph{can} be made, generally, and correspond to different divisions between mean flow and fluctuations, typically important at quadratic order in fluctuation amplitude. The issues are discussed in \cite{SoRo10} and \cite{GiVa18}, with references to related literature. %Naturally in any application some choice has to be made, as in section \ref{ssecbragex}. 

For the Braginsky dynamo, we gave a sketch of the classic dynamo set-up  in a general geometry. Here there is a background flow giving an $\omega$-effect, converting transverse field components $B_1$, $B_2$ to azimuthal field $B_3$, but there is no feedback from the background geometry or flow that can regenerate the transverse field, even with diffusion $\eta>0$. In short, we restricted attention to a limited, but still wide, family of possible metrics $g(x,y)$ and background flows $\ub(x,y)$. We then introduced some waves or fluctuations, giving $z$-dependence by means of a Lagrangian map $\xi$ and showed how the pull back of the equations under $\xi$ preserves the structure of the ideal terms, while the diffusion term, present for $\eta>0$, can be understood in terms of the pull back of the codifferential operator $\delta$. Averaging over a family of such waves, we then obtained the $\alpha$-effect, written in various  forms in (\ref{eqfullalpha8b}), (\ref{eqalphafinal1}--\ref{eqalphafinal4}) and (\ref{eqalphapert1}, \ref{eqalphapert2}, \ref{eqfullalpha3a}), that can close the dynamo loop and lead to a sustained or growing field. 

We stress that our study is a sketch, aiming at exposing the geometry behind the origin and definition of the $\alpha$-tensor. 
Nonetheless the approach using the compact notation and identities of differential geometry allows us to interpret the $\alpha$-effect in terms of pull backs and connections, derive established formulae and give new expressions, all valid for arbitrary coordinate systems. We note though that writing down  formulae for the $\alpha$ tensor is just one element of the bigger picture, which requires the scaling of field component magnitudes in terms of the magnetic Reynolds number, setting up a suitable eigenvalue problem, and keeping track of the order of the errors involved. However this is already well studied in the literature and we refer the reader to \cite{SoRo14} for detailed discussion. 

%\AG{
%Potentially to do:
%
%* any link from alpha to heliicity, or bounds involving helicity?
%
%* ? send to Arter, Oliver?
%
%* any-effects from a pull-back and viscosity that parallel the discussion here (AKA-effect or similar)?
%
%* look out for connections to $\alpha$ calculations in \cite{SoRo14} --- I had some trouble linking theirs to mine - they seem to have yet another version of the alpha-effect!
%
%}

\section*{Acknowledgements}

ADG is very grateful to the Leverhulme Trust for the award of a Research Fellowship, which enabled the continuing collaboration with JV and the work presented in this paper. AG also benefitted from STFC funding, research grant ST/R000891/1. We are grateful to Darryl Holm, Marcel Oliver and Andrew Soward for many useful discussions, and to the referees for constructive criticisms and additional  references. We thank Leo Stein for the use of his notes in \cite{St17}.

\appendix

\section{Derivation of (\ref{eqdeltanabla})} \label{appdivergence}

We establish (\ref{eqdeltanabla}) in three dimensions, using the coordinate expression
\begin{equation}
(\star B)_i = \tfrac{1}{2} g_{ij} \mu^{jkl} B_{kl}
\end{equation}
for the Hodge $\star$ operator applied to a 2-form $B$  (see (\ref{eqexplicithodgestar})). 
Noting (\ref{eqdcompcov}), we have 
\begin{align}
(\delta B)_i = ( {\star}\dd{\star}B)_i & = g_{ij} \,\mu^{jkl}  \,\nabla_{k}(g_{lm} \, \tfrac{1}{2} \,\mu^{mnp} \, B_{np})  ,\\
& = g_{ij} \,\mu^{jkl}  \, g_{lm} \, \tfrac{1}{2} \,\mu^{mnp} \, \nabla_{k} B_{np}  \\
& = g^{pk}\,  \nabla_k B_{ip} , 
\end{align}
using that
\begin{equation}
\mu^{jkl} \,  \,g_{lm}\, \mu^{mnp} = g^{jn} \, g^{pk} - g^{jp}\,  g^{nk} = 2 \, g^{j[n}\,  g^{p]k}, 
\label{eqggmumuid}
\end{equation} 
 to obtain the last line. This gives $\delta B$ as minus the \emph{divergence} of the 2-form $B$ as defined by \cite{Fr97}.

\section{The $\alpha$-effect from helical waves} \label{appbragex}

As a basic example of the calculation of $\alpha$ in sections \ref{ssecmotivation} and \ref{ssecdiscussalpha}, and to link with well-established theory, suppose we take a situation similar to that in section \ref{ssecalfvenwave}, with $g = \dd x^2 + \dd y^2 + \dd z^2$, and some waves given by Lagrangian displacements of the form 
\begin{equation}
\xi (x,y,z,t) = ( x + F(z-ct), y + H(z-ct), z). 
\end{equation}
Using (\ref{eqpullbackd}), the pulled back metric and its inverse are 
\begin{equation}
\gh_{ij} =   \begin{pmatrix} \;\; 1 \;\; & \;\; 0\;\;  & F'  \\ 0  & 1 & H' \\  F' & H' & 1 + F^{\prime 2} + H^{\prime 2} \end{pmatrix} , \quad
\gh^{ij} =   \begin{pmatrix} \;\; 1+ F^{\prime2}  \;\; & \;\; H' F' \;\;  & - F'  \\ H'F'  & 1+H^{\prime2}  & - H' \\  - F' & - H' & 1 \end{pmatrix} , 
\label{eqmetrics}
\end{equation}
and have determinant $1$ so that $\varsigmah = 1$ also (the map $\xi$ is volume-preserving). Here we find from (\ref{eqalpha3}) that
\begin{equation}
\alpha = \eta \, \langle  \gh_{31} \upartial_z \gh_{23} - \gh_{32} \upartial_z \gh_{13}\rangle = \eta \, \langle F'H'' - H' F''\rangle. 
\end{equation}

Thus for example if we take a case with no background flow $\ub = 0$ and a wave with 
\begin{equation}
u = w = U \sin ( k (z-ct)) \, \upartial_x + U \cos(k(z-ct)) \, \upartial_y, 
\end{equation}
then we have the fluctuating map given by 
\begin{equation}
F(s) = U \omega^{-1} \cos ks, \quad 
H(s) = - U \omega^{-1} \sin ks, \quad 
c \equiv \omega / k , 
\end{equation}
and %from (\ref{eqalpha3}),
\begin{equation}
\alpha = - \eta U^2 k^3 \omega^{-2} .
\label{eqalphahelical1}
\end{equation}
Note that the (kinetic) helicity of the wave is given by the integral of the helicity form defined by
\begin{equation}
h_\textrm{K}   = \nu \wedge \dd\nu = - k^{-2} \omega^{2}\, \langle F' H'' - H' F'' \rangle \, \dd x \wedge \dd y \wedge \dd z = U^2 k \, \dd x\wedge \dd y \wedge \dd z, 
\end{equation}
and so here positive helicity gives rise to a negative cofficient, $\alpha = - \eta k^2 \omega^{-2} h_\textrm{K}$; this is in keeping with the traditional sign convention in the literature. 
%\AG{see HKM book 7.69, 7.77 and later sections.} 
In this example there is no background flow and so the displacements given by $\xi$ are easily related to the flow field, with  $u=w$. If there is a background flow then this needs to be taken into account via (\ref{eqpullbacku}). 

%Naturally, in obtaining a complete $\alpha\omega$-dynamo we need a mean flow $\ub = u_3(x,y)\, \upartial_z$ with some non-trivial dependence on $(x,y)$ to provide the $\omega$-effect; these factors need to be taken into account in more complete models, but our goal here is only to isolate the individual effects, not to create a complete and consistent $\alpha\omega$-dynamo. 

%Turning to the more general $\alpha$-effect formula (\ref{eqfullalpha3a}), we have $\alpha_3^{12} = - \alpha_3^{21} = \tfrac{1}{2} \alpha$ as given in (\ref{eqalphahelical1}). For the given metric (\ref{eqmetrics}) the only other terms that could be non-zero in fact vanish under $z$-averaging, namely 
%%
%\begin{equation}
%\alpha_1^{12} = - \alpha_1^{21} = \tfrac{1}{2} \eta \langle H'' \rangle = 0 , \quad
%\alpha_2^{12} = - \alpha_2^{21} = - \tfrac{1}{2} \eta \langle F'' \rangle  = 0 . 
%\label{eqalphacoeffs} 
%\end{equation}
%%

%\JVcom{There's a problem with the above para in that it uses a formula from Appendix C that hasn't appeared yet. I think the para can be removed with the fix suggested for the para below.}

As a check it is %also 
interesting to calculate the $\alpha$ tensor components from (\ref{eqfullalpha8b}). The Christoffel symbols $\Gamma^i_{jk}$ for $g$ are all zero and so we compute the $\tilde{\Gamma}^i_{jk}$ for $\gh$. We have then using standard notation 
\begin{equation}
\Gammah_{133} = F'', \quad \Gammah_{233} = H'', \quad \Gammah_{333} = F'F'' + H'H'', \quad C^1_{33} =  \Gammah^1_{33} = F'', \quad C^2_{33} = \Gammah^2_{33} = H'', 
\end{equation}
other terms being zero.
% This gives the only non-zero terms for $\alpha_i^{jk}$ as in (\ref{eqalphacoeffs}) and just above. 
This gives the only non-zero terms for $\alpha_i^{jk}$ as $\alpha_3^{12} = - \alpha_3^{21} = \tfrac{1}{2} \alpha$ with $\alpha$ given in (\ref{eqalphahelical1}). 
Referring to the calculations of section \ref{ssecpertalpha}, in this example $\eps q^1=\eps q_1 = F$, $\eps q^2= \eps q_2 =H$ and there is agreement between the formulae for the $\alpha$ tensor here and in that section. 
%\JVcom{Replace the sentence by:

\section{Explicit calculations of the $\alpha$-effect} \label{appexplicitalpha}

In this appendix we relate our calculations of the $\alpha$-effect to formulae found in the literature.
To do so, we first give an expression for the $\alpha$ tensor alternative to (\ref{eqfullalpha8b}) and closer to those found elsewhere. This formula is obtained from (\ref{eqfulldiff}) rewritten as
\begin{equation}
\emfh_i   = 
 -  \eta \,  \gh_{ij} \,\muh^{jkl}  \,\nabla_{k}(\gh_{lm} \, \tfrac{1}{2} \,\muh^{mnp} \, \bfluxh_{np}) , 
  \label{emfagain1}
\end{equation}
making the effect of the pull back on the metric and volume form explicit. %, and slipping in an average over some family of waves.
Here we have used the pulled back volume form $\muh = \varsigmah^{-1} \dd x^1\wedge \dd x^2 \wedge \dd x^3$ with covariant components $\muh_{ijk} = \varsigmah^{-1} \eps_{ijk}$ and the corresponding contravariant tensor $\muh^{ijk} = \varsigmah \eps^{ijk}$, together with (\ref{eqdcompcov}).
%We have also used the fact that in the definition of $d$ acting on a 1-form in (\ref{eqdcomp}) partial derivatives may be replaced by covariant derivatives (given that there is no torsion for a covariant derivative induced by a metric).  
Comparing (\ref{emfagain1}) with (\ref{eqalphagamma}) gives the alternative to (\ref{eqfullalpha8b}--\ref{eqgammagamma1}),
\begin{align}
\alpha_i ^{np} & = - \eta \, \langle \tfrac{1}{2}\,  \gh_{ij} \,\muh^{jkl} \,\nabla_{k}(\gh_{lm}\, \muh^{mnp}  ) \rangle, 
\label{eqfullalpha3a} \\
\gamma_i^{knp} & = - \eta \,  \langle \tfrac{1}{2} \, \gh_{ij}\, \muh^{jkl} \,  \, \gh_{lm}\, \muh^{mnp}  \rangle , \label{eqgamma3b}
\end{align}

% This $\alpha$ tensor gives all the coupling terms for any metric. In the differential geometry framework, the actual tensor is $\alpha = \alpha_i^{np} \, \dd x^i \otimes \upartial_n \otimes \upartial_p$ and can be thought of as a map from 2-forms to 1-forms (a \emph{1-form valued 2-vector}): a 2-form such as the magnetic flux $\bfluxh_{np}$ is contracted on the second \emph{leg} of the tensor to give the $\alpha$-effect piece, $\alpha_i^{np} \bfluxh_{np} \, \dd x^i $, of the 1-form $\mathcal{E}$. 

It is instructive to show that (\ref{eqfullalpha3a}, \ref{eqgamma3b}) are equivalent to (\ref{eqfullalpha8b}, \ref{eqgammagamma1}) in a direct manner. For $\gamma$, this is immediate using (\ref{eqggmumuid}). For $\alpha$, we compute
\begin{align}
-   \gh_{ij} \,\muh^{jkl} \,\nabla_{k}(\gh_{lm}\, \muh^{mnp}  ) &= \gh_{ij} \,\muh^{jkl} \, (\nablah_k-\nabla_k) (\gh_{lm}\, \muh^{mnp}  ) \\
&= \gh_{ij} \,\muh^{jkl}  \left( C_{ks}^n  \, \gh_{lm} \muh^{msp} + C_{ks}^p \, \gh_{lm} \muh^{mns} - C_{kl}^s  \, \gh_{sm} \muh^{mnp} \right),
\end{align}
using that $\nablah \gh = 0$ and $\nablah \muh=0$ and (\ref{eqconnection}). The last term vanishes by (anti)symmetry. From (\ref{eqggmumuid}) we then find
\begin{equation}
-   \gh_{ij} \,\muh^{jkl} \,\nabla_{k}(\gh_{lm}\, \muh^{mnp}  ) = \gh^{kp} C_{ki}^n - \gh^{ks} C_{ks}^n \delta_i^p + \gh^{ks} C_{ks}^p \delta_i^n - C_{ki}^p \gh^{kn} = 2 \gh^{k[p} C_{ki}^{n]} + 2  \gh^{ks} C_{ks}^{[p} \delta_i^{n]},
\end{equation}
and using this in (\ref{eqfullalpha3a}) recovers (\ref{eqfullalpha8b}), as expected.

%Similar remarks apply to the effective diffusion tensor $\gamma_i^{knp}$ which can be simplified; a short calculation shows that 
%%
%\begin{equation}
%\muh^{jkl} \,  \, \gh_{lm}\, \muh^{mnp} = \gh^{jn} \, \gh^{pk} - \gh^{jp}\,  \gh^{nk} = 2 \, \gh^{j[n}\,  \gh^{p]k}, 
%\label{eqggmumuid}
%\end{equation}
%%
%and with this we have
%%
%\begin{equation}
%\gamma_i^{knp} = - \eta \,\delta_i^{[n} \, \langle \gh^{p]k} \rangle , \quad
%\gamma_i^{knp}\, \nabla_k  \bfluxh_{np}  = - \eta \, \langle \gh^{pk} \rangle \,\nabla_k  \bfluxh_{ip} . 
%\end{equation}
%%
%Thus the effective diffusion tensor involves the average of the pulled-back metric $\gh$ over the ensemble of waves. 

We now turn to concrete calculations, recalling that in each realisation of one of our flows, the fluctuating map is  $x \to \xi(x)$, so if the point with coordinates $x^i$ marks a Lagrangian parcel in the background flow at some time, $\xi^i(x)$ are its coordinates in the full flow. The pull back from $\xi(x)$ to $x$ for vectors and 1-forms is simply the Cauchy solution, namely
\begin{equation}
\vh(x)^i = \frac{\upartial x^i}{\upartial \xi^j}\,  v(\xi)^j , \quad
\sigmah(x)_i =  \frac{\upartial \xi^j}{\upartial x^i}\,  \sigma(\xi)_j .
\end{equation}
We also need the pull back of the metric, which is a twice covariant tensor $g = g_{ij}\,  \dd x^i \otimes \dd x^j$. This is given by
\begin{equation}
\gh(x)_{ij} = [(\xi^*g )(x)]_{ij}  =  \frac{\upartial \xi^k}{\upartial {x}^i}  \, \frac{\upartial \xi^l}{\upartial {x}^j}    \, g_{kl}(\xi).
\end{equation}
Correspondingly, $\varsigmah$ defined by $\varsigmah(x)^{-2}= \det \tilde g(x)$ is given by
\begin{equation}
\varsigmah(x)^{-1}  =  \varsigma^{-1}(\xi) \det \left(  \frac{\upartial \xi^i}{\upartial {x}^j} \right), 
\label{eqvarsigmatrans} 
\end{equation}
Substituting these into (\ref{eqfullalpha3a}) and using $\muh^{ijk} = \varsigmah\,  \eps^{ijk} $ gives an explicit formula for the $\alpha$ tensor in terms of derivatives of the map $\xi$. Note that we are not assuming incompressible flow here; if we do then $\varsigmah = \varsigma$ and (\ref{eqvarsigmatrans}) becomes the condition that the map $\xi$ must satisfy to be volume preserving (assuming the actual flow and background flow are incompressible also).

%Note that if we have incompressible flows so that $\phi$, $\phib$, $\xi$ are all volume preserving and so $\varsigma = \varsigmah$ (this quantity could be a function of coordinates $(x,y)$ note, for example if we are using cylindrical or spherical polar coordinates $(x,y,z) = (r,\theta, z) $ or $(r, \theta, \phi)$). In this case we have the condition that $\xi$ be volume-preserving, i.e.\ the volume form $\mu = \varsigma^{-1} \dd x \wedge dy \wedge \dd z$ is invariant under pull back, so that the relation 
%%
%\begin{equation}
%\varsigmah(\xh)^{-1}  = \varsigma(x)^{-1} \det \left(  \frac{\upartial x^i}{\upartial \xh^j} \right) 
%\end{equation}
%%
%holds. Substituting these into (\ref{eqfullalpha2a}) enables calculations to be undertaken for any choice of underlying metric $g$, and fluctuations on a background flow. 

To relate this to formulae in the literature, suppose that we are in Euclidean space with the underlying metric $g = \dd x^2 + \dd y^2 + \dd z^2$ and so $\mu = \dd x \wedge \dd y \wedge \dd z$, $\varsigma=1$, and allow the maps $\xi$ to be compressible. Then the $\alpha$-effect formula (\ref{eqfullalpha3a}) becomes 
\begin{equation}
\alpha_i ^{np} =   - \eta\, 
\left\langle
\frac{\upartial \xi^q}{\upartial x^i} \,
\frac{\upartial \xi^q}{\upartial  x^j} 
\,\varsigmah\, \eps^{jkl} \,
\frac{\upartial}{\upartial x^k} 
\left(
\frac{\upartial \xi^r}{\upartial x^l} \,
\frac{\upartial \xi^r}{\upartial x^m} 
\, \varsigmah 
\right)  \,\tfrac{1}{2} \, \eps^{mnp}  \right\rangle, 
\label{eqfullalpha1b}
\end{equation}
with 
\begin{equation}
\varsigmah(x) =  \det \left(  \frac{\upartial x^i}{\upartial \xi^j} \right) . 
\label{eqvarsigmadet}
\end{equation}
In (\ref{eqfullalpha1b}) the term ${\upartial \xi^r}/{\upartial x^l} $ can come out of the differentation with respect to $x^k$ and from (\ref{eqvarsigmadet}) it follows that 
\begin{equation}
\eps^{jkl}\, \frac{\upartial \xi^q}{\upartial x^j} \,\frac{\upartial \xi^r}{\upartial x^l} = \varsigmah^{-1} \, \eps^{qsr}\, \frac{\upartial x^k}{\upartial \xi^s} \, , 
\end{equation}
so that (\ref{eqfullalpha1b}) becomes
\begin{equation}
\alpha_i ^{np} =   - \eta\, 
\left\langle
\frac{\upartial \xi^q}{\upartial x^i} \,
 \eps^{qsr} \,
\frac{\upartial}{\upartial \xi^s} \,
\left(
\frac{\upartial \xi^r}{\upartial x^m} 
\, \varsigmah 
  \,\tfrac{1}{2} \, \eps^{mnp}    \right) \right\rangle . 
\label{eqfullalpha2b}
\end{equation}
Both equations (\ref{eqfullalpha1b}) and (\ref{eqfullalpha2b}) are given in \cite{RoSo06a,RoSo06b} (with $\varsigmah \to \mathcal{J}^{-1}$, $\xi^i \to x^*_i$, $x^i \to x_i$, to map our notation to theirs). 
%
%
%\JVcom{I'm not sure that the following paragraph is useful. It seems to sketch the computations carried out more carefully in \S4.3.}
%\AG{Perhaps it isn't so helpful.... but.... my goal was to show where the equations  (\ref{eqfullalpha1b}) and (\ref{eqfullalpha2b}) actually come from --- in other words all the manipulations people do, e.,g. to write down these two versions, just correspond to the pull back and push forwards taking place at different points of the calculations.... I'm inclined to leave the paragraph --- it is not so useful for us, but my goal is to make contact with formulae elsewhere. }

%\JVcom{What follows is rather cryptic and could be omitted.}

While the formulae (\ref{eqfullalpha1b}) and (\ref{eqfullalpha2b}) may be derived without reference to the framework we present, by following standard rules of multivariate calculus, they simply correspond to applying pull backs and push forwards at different points of the same calculation. We now illustrate this, 
% transformations that give equivalent versions of the $\alpha$-effect have analogues in the differential geometric framework, 
noting that notation here becomes awkward.  
Let us temporarily write $g$ as an operator to give $g \bvec = \bvec_{\flat}$, and $\mu$ as an operator with $\mu B = b$ for $b\ip \mu = B$, and likewise for $\gh$ and $\muh$. Then we can expand the diffusion operator $- \nabla^2$ as 
\begin{equation}
\dd {\star} \dd{\star} \bflux = \dd g \mu \dd g \mu \bflux. 
\label{eqexpand1}
\end{equation}
We set $\bfluxh = \xi^* \bflux$, $\bflux = \xi_* \bfluxh$, and apply $\xi^*$ to (\ref{eqexpand1}); The term we then have in our pulled back induction equation can be written as 
\begin{align}
\xi^* (\dd g \mu \dd g \mu \,\bflux) & = \xi^* (\dd g \mu \dd g \mu \,\xi_*\bfluxh) \\
& =\dd \gh \muh \dd \gh \muh \bfluxh , 
\end{align}
corresponding to the structure (\ref{eqfullalpha1b}) for the $\alpha$-effect, or as
\begin{align}
\xi^* (\dd g \mu \dd g \mu \,\bflux) & = \dd \xi^* [ g \mu \dd g  \,\xi_*( \muh \bfluxh)], 
\end{align}
in line with (\ref{eqfullalpha2b}). There is flexibility to undertake operations at different points on $\Mc$, and this is clarified by the language of pull backs and push forwards, albeit that it is not easy to write out cleanly in this calculation. On the other hand the differential geometric setting does avoid explicit repeated use of the chain rule and properties of determinants; these are built in.

\end{document}